\documentclass[aps,showkeys,showpacs,twocolumn,showpacs,preprintnumbers,amsmath,amssymb,superscriptaddress,floatfix,nofootinbib]{revtex4}
\usepackage{graphicx,color,dcolumn,booktabs,bm}
\usepackage{longtable,lscape}
\usepackage{txfonts}
\usepackage{overpic}
\usepackage{epsfig}
\usepackage{amssymb}
\usepackage{rotating}
\usepackage{epstopdf}
\usepackage{indentfirst}
\usepackage{feynmf}   
\usepackage{slashed}  
\usepackage{cases}
\usepackage{color}
\usepackage{multirow}
\usepackage{graphicx,color,dcolumn,booktabs,bm}
\usepackage{cases}
\usepackage{array}

\graphicspath{{Figures/}} %

\begin{document}

\title{Strong decays of the higher excited $\Lambda_Q$ and $\Sigma_Q$ baryons }

\author{Qi-Fang L\"{u}} \email{lvqifang@hunnu.edu.cn}
\affiliation{  Department
of Physics, Hunan Normal University,  Changsha 410081, China }

\affiliation{ Synergetic Innovation
Center for Quantum Effects and Applications (SICQEA), Changsha 410081,China}

\affiliation{  Key Laboratory of
Low-Dimensional Quantum Structures and Quantum Control of Ministry
of Education, Changsha 410081, China}
\author{Xian-Hui Zhong} \email{zhongxh@hunnu.edu.cn}
\affiliation{  Department
of Physics, Hunan Normal University,  Changsha 410081, China }

\affiliation{ Synergetic Innovation
Center for Quantum Effects and Applications (SICQEA), Changsha 410081,China}

\affiliation{  Key Laboratory of
Low-Dimensional Quantum Structures and Quantum Control of Ministry
of Education, Changsha 410081, China}

\begin{abstract}

 In this work, we preform a systematic study of the strong decays for the higher singly heavy baryon resonances $\Lambda_Q(3S)$, $\Lambda_Q(2P)$, $\Lambda_Q(2D)$, $\Lambda_Q(1F)$, $\Sigma_Q(3S)$, $\Sigma_Q(2P)$, $\Sigma_Q(2D)$, and $\Sigma_Q(1F)$ within the $^3P_0$ model. Our results show that most of the $\lambda-$mode higher excited $\Lambda_Q$ and $\Sigma_Q$ states have a relatively narrow width, and mainly decay into two-body final states with one heavy meson plus one light baryon. Our calculations provide abundant theoretical information and demonstrate the general feature for these higher states, which may be valuable for future experimental searches and establishing the singly heavy baryon spectrum.
\end{abstract}


\maketitle

\section{Introduction}{\label{introduction}}

Recently, the observations of the singly heavy baryon spectrum have achieved
significant progress in experiments. In 2017, a new charmed resonance $\Lambda_c(2860)$ was observed in the $D^0p$ final state by the LHCb Collaboration~\cite{Aaij:2017vbw}. Just in the same year, they also found five narrow $\Omega_c$ resonances in the $\Xi_c K$ invariant mass~\cite{Aaij:2017nav}, and most of them were confirmed by the Belle Collaboration subsequently~\cite{Yelton:2017qxg}. In 2018, two bottom resonances $\Sigma_b(6097)$ and $\Xi_b(6227)$ were found in the $\Lambda_b \pi$ and $\Lambda_b K$ final states at LHCb, respectively~\cite{Aaij:2018tnn,Aaij:2018yqz}. In 2019, the LHCb Collaboration reported the observation of two bottom baryon resonances $\Lambda_b(6146)$ and $\Lambda_b(6152)$ in the $\Lambda_b \pi^+ \pi^-$ channel, which are good candidates of the $D-$wave $\Lambda_b$ doublet~\cite{Aaij:2019amv}. The experimental progress provides fruitful information for us to
establish the low-lying singly heavy baryon spectrum.

\begin{figure}[!htbp]
\includegraphics[scale=0.65]{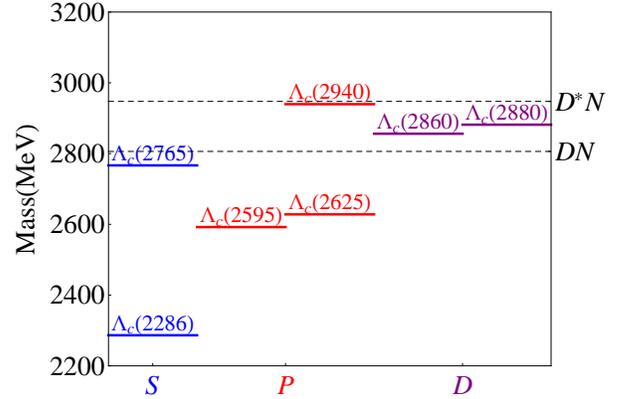}
\vspace{0.0cm} \caption{The mass spectrum of the $\lambda-$mode $\Lambda_c$ family.}
\label{mass}
\end{figure}

Focusing on the nonstrange sectors $\Lambda_Q$ and $\Sigma_Q$ ($Q=c,b$), plenty of theoretical works have been made to investigate these low-lying singly heavy baryons~\cite{Capstick:1986bm,Ebert:2007nw,Ebert:2011kk,Chen:2014nyo,Yoshida:2015tia,Chen:2016iyi,Shah:2016mig,Thakkar:2016dna,Cheng:2006dk,Valcarce:2008dr,
Roberts:2007ni,Garcilazo:2007eh,Karliner:2015ema,Mao:2015gya,Chen:2018vuc,Wang:2018fjm,Yang:2018lzg,Aliev:2018vye,Cui:2019dzj,Jia:2019bkr,Nagahiro:2016nsx,
Huang:1995ke,Albertus:2005zy,Zhong:2007gp,Hussain:1999sp,Ivanov:1999bk,Guo:2007qu,Wang:2019uaj}. In the traditional quark model, the $\Lambda_Q$ and $\Sigma_Q$ states belong to the antisymmetric flavor structure $\bar 3_F$ and symmetric flavor structure $6_F$, respectively. Take the low-lying $\Lambda_c$ spectrum for example, the $\Lambda_c(2286)$ is the ground $\Lambda_c$ state, and the $\Lambda_c(2595)$ and $\Lambda_c(2625)$ can be explained as the two $P-$wave states~\cite{Chen:2017sci,Wang:2017kfr,Arifi:2018yhr,Guo:2019ytq}. The $\Lambda_c(2765)$ can be interpreted as the $\Lambda_c(2S)$ state, whose isospin has been confirmed to be zero by Belle Collaboration~\cite{Abdesselam:2019bfp}. The $\Lambda_c(2860)$ and $\Lambda_c(2880)$ resonances are consistent with the theoretical predictions of the $\Lambda_c(1D)$ doublet with $J^P=3/2^+$ and $J^P=5/2^+$, respectively~\cite{Chen:2017aqm,Wang:2017vtv,Yao:2018jmc}. Under this conventional interpretation, the low-lying $\lambda-$mode $\Lambda_c$ spectrum with $N=0,1$ and 2 shells has been established, and is presented in Fig.~\ref{mass}. Other theoretical investigations of the low-lying $\Sigma_c$, $\Lambda_b$, and $\Sigma_b$ spectra also achieve great successes. More interpretations and detailed discussions of these singly heavy baryons can be found in Refs.~\cite{Chen:2016spr,Crede:2013sze,Cheng:2015iom,Richard:1992uk,Klempt:2009pi}.

Above the $DN$ threshold, three $\Lambda_Q$ resonances $\Lambda_c(2860)$, $\Lambda_c(2880)$, and $\Lambda_c(2940)$ have been found in experiments. Although the $\Lambda_c(2860)$ and $\Lambda_c(2880)$ lie above the $DN$ threshold, their partial widths decaying into the $DN$ channel are small for the limited phase space. The $\Lambda_c(2940)$ resonance may favor the $\lambda-$mode $J^P = 3/2^-$ $\Lambda_c(2P)$ state in the conventional quark model~\cite{Aaij:2017vbw,Ebert:2011kk,Chen:2014nyo,Lu:2018utx}, which is the only higher excited $\Lambda_c$ state in the $N=3$ shell. Unlike $\Lambda_c(2860)$ and $\Lambda_c(2880)$, the $\Lambda_c(2940)$ has large partial decay width of $DN$ decay mode~\cite{Lu:2018utx}.

It can be noticed that the dominant decay modes of low-lying singly heavy baryons are heavy baryon plus light meson channels. While for the high-lying states, the heavy meson plus light baryon channels, such as $D^{(*)}N$, $D^{(*)}\Delta$, $B^{(*)}N$, and $B^{(*)}\Delta$, are open, and these channels may be the dominant decay modes. Especially, the $\rho-$mode $\Lambda_Q$ and $\Sigma_Q$ states decaying into the heavy meson plus light baryon final states should be highly suppressed since the light quark system can be regarded as the spectator. If one observes a new resonance in $D^{(*)}N$, $D^{(*)}\Delta$, $B^{(*)}N$, or $B^{(*)}\Delta$ channel, this resonance strongly favors the $\lambda-$mode assignment.

Except for the $\Lambda_c(2940)$, the theoretical attentions are not enough for the higher $\Lambda_Q$ and $\Sigma_Q$ excitations with $N=3$ and 4 shells. There are few investigations of the strong decays for these higher $\Lambda_Q$ and $\Sigma_Q$ resonances, although some quark model studies of the mass spectra can be found in the literature~\cite{Capstick:1986bm,Ebert:2007nw,Ebert:2011kk,Chen:2014nyo,Yoshida:2015tia,Chen:2016iyi,Shah:2016mig,Thakkar:2016dna}. In fact, the decay properties of the higher $\Lambda_Q$ and $\Sigma_Q$ excitations are crucial for our searching for them in experiments.  If the higher resonances dominantly decay into the channels with one heavy meson plus one light baryon, they may be relatively easy to be observed in these special channels by future LHCb and BelleII experiments just as the $\Lambda_c(2940)$ resonance has been observed in the $D^0p$ final state. In this work, we study the strong decays of the higher excited $\Lambda_Q(3S)$, $\Lambda_Q(2P)$, $\Lambda_Q(2D)$, $\Lambda_Q(1F)$, $\Sigma_Q(3S)$, $\Sigma_Q(2P)$, $\Sigma_Q(2D)$, and $\Sigma_Q(1F)$ states within the $^3P_0$ quark pair creation model. Our results indicate that most of the $\lambda-$mode higher excited $\Lambda_Q$ and $\Sigma_Q$ states have relatively narrow total widths, and mainly decay into the heavy meson plus light baryon final states. Our predictions show the general feature and provide helpful theoretical information for these higher states, which may be valuable for future experimental searches.

This paper is organized as follows. The formulas of the quark pair creation model and adopted notations are briefly introduced in Sec.~\ref{model}. The strong decay behaviors of the higher excited $\Lambda_Q$ and $\Sigma_Q$ baryons are estimated in Sec.~\ref{decay1} and Sec.~\ref{decay2}, respectively. A short summary is presented in the last section.

\section{$^3P_0$ Model}{\label{model}}

In present work, the $^3P_0$ model is adopted to estimate the OZI-allowed two-body strong decays of the higher excited $\Lambda_Q$ and $\Sigma_Q$ baryons. This model has been widely employed to investigate the strong decays of conventional hadrons and meet with considerable successes~\cite{micu,3p0model1,3p0model2,3p0model5,Chen:2007xf,Segovia:2012cd,Ferretti:2014xqa,Mu:2014iaa,Godfrey:2015dva,Zhao:2017fov,Liang:2019aag}. For a heavy baryon $A$, it can decay into two final states $B$ and $C$ via a $J^{PC}=0^{++}$ quark-antiquark pair~\cite{micu}, and the transition operator can be taken as
\begin{eqnarray}
T&=&-3\gamma\sum_m\langle 1m1-m|00\rangle\int
d^3\boldsymbol{p}_4d^3\boldsymbol{p}_5\delta^3(\boldsymbol{p}_4+\boldsymbol{p}_5)\nonumber\\
&&\times {\mathcal{Y}}^m_1\left(\frac{\boldsymbol{p}_4-\boldsymbol{p}_5}{2}\right
)\chi^{45}_{1,-m}\phi^{45}_0\omega^{45}_0b^\dagger_{4i}(\boldsymbol{p}_4)d^\dagger_{4j}(\boldsymbol{p}_5),
\end{eqnarray}
where $\gamma$ is a dimensionless constant of the $q_4\bar{q}_5$ pair-production strength. $\boldsymbol{p}_4$ and
$\boldsymbol{p}_5$ are the momenta of the created quark $q_4$ and
antiquark $\bar{q}_5$, respectively. The $P-$wave momentum-space distribution of the $q_4\bar{q}_5$ can be described by the solid harmonic polynomial
${\cal{Y}}^m_1(\boldsymbol{p})\equiv|p|Y^m_1(\theta_p, \phi_p)$. $\chi_{{1,-m}}^{45}$, $\phi^{45}_{0}=(u\bar u + d\bar d +s\bar s)/\sqrt{3}$, and
$\omega^{45}=\delta_{ij}$ are the spin triplet, flavor singlet, and color singlet wave functions of the $q_4\bar{q}_5$, respectively. The $b^\dagger_{4i}(\boldsymbol{p}_4)d^\dagger_{4j}(\boldsymbol{p}_5)$ stands for the creation operators, where the $i$ and $j$ are the color indices.

The definitions of the mock states are adopted. For instance, the total wave function of initial baryon $A$ can be taken as~\cite{Hayne:1981zy}
\begin{eqnarray}
&&|A(n^{2S_A+1}_AL_{A}\,\mbox{}_{J_A M_{J_A}})(\boldsymbol{P}_A)\rangle
\equiv \nonumber\\
&& \sqrt{2E_A}\sum_{M_{L_A},M_{S_A}}\langle L_A M_{L_A} S_A
M_{S_A}|J_A
M_{J_A}\rangle \int d^3\boldsymbol{p}_1d^3\boldsymbol{p}_2d^3\boldsymbol{p}_3\nonumber\\
&&\times \delta^3(\boldsymbol{p}_1+\boldsymbol{p}_2+\boldsymbol{p}_3-\boldsymbol{P}_A)\psi_{n_AL_AM_{L_A}}(\boldsymbol{p}_1,\boldsymbol{p}_2,\boldsymbol{p}_3)\chi^{123}_{S_AM_{S_A}}
\phi^{123}_A\omega^{123}_A\nonumber\\
&&\times  \left|q_1(\boldsymbol{p}_1)q_2(\boldsymbol{p}_2)q_3(\boldsymbol{p}_3)\right\rangle,
\end{eqnarray}
which satisfies the normalization condition
\begin{eqnarray}
\langle A(\boldsymbol{P}_A)|A(\boldsymbol{P}^\prime_A)\rangle=2E_A\delta^3(\boldsymbol{P}_A-\boldsymbol{P}^\prime_A).
\end{eqnarray}
The $\boldsymbol{P}_A$, $\boldsymbol{p}_1$, $\boldsymbol{p}_2$, and $\boldsymbol{p}_3$ are the momenta of the baryon $A$, quark $q_1$, quark $q_2$, and quark $q_3$, respectively. $E_A$ is the total energy of the baryon $A$. $\chi^{123}_{S_AM_{S_A}}$, $\phi^{123}_A$, $\omega^{123}_A$, $\psi_{n_AL_AM_{L_A}}(\boldsymbol{p}_1,\boldsymbol{p}_2,\boldsymbol{p}_3)$ are the spin, flavor, color, and
space wave functions, respectively. The definitions final states $B$ and $C$ are similar to that of initial state $A$, which can be found in Ref.~\cite{Chen:2007xf}.

\begin{figure}[!htbp]
\includegraphics[scale=0.78]{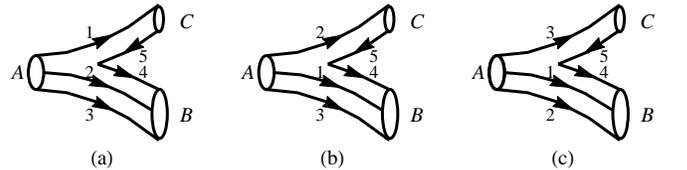}
\vspace{0.0cm} \caption{The baryon decay process $A\to B+C$ in the $^3P_0$ model.}
\label{qpc}
\end{figure}

For the decay of the singly heavy baryons $\Lambda_Q$ and $\Sigma_Q$, three possible rearrangements exist
\begin{eqnarray}
A(q_1,q_2,Q_3)+P(q_4,\bar q_5)\to B(q_2,q_4,Q_3)+C(q_1,\bar q_5),\\
A(q_1,q_2,Q_3)+P(q_4,\bar q_5)\to B(q_1,q_4,Q_3)+C(q_2,\bar q_5),\\
A(q_1,q_2,Q_3)+P(q_4,\bar q_5)\to B(q_1,q_2,q_4)+C(Q_3,\bar q_5),
\end{eqnarray}
which are also performed in Fig.~\ref{qpc}. It can be noticed that he first and second cases correspond to the heavy baryon plus the light meson channels, while the third one denotes the light baryon plus the heavy meson decay mode.

The helicity amplitude ${\cal{M}}^{M_{J_A}M_{J_B}M_{J_C}}$ of the decay process $A\to B+C$ can be related to the $S$ matrix
\begin{eqnarray}
\langle
f|S|i\rangle=I-i2\pi\delta(E_f-E_i){\cal{M}}^{M_{J_A}M_{J_B}M_{J_C}}.
\end{eqnarray}
Take the $A(q_1,q_2,b_3)+P(q_4,\bar q_5)\to B(q_1,q_4,b_3)+C(q_2,\bar q_5)$ shown in Fig. 1(b) as an example, the helicity amplitude ${\cal{M}}^{M_{J_A}M_{J_B}M_{J_C}}$ can be expressed as,
\begin{eqnarray}
&&\delta^3(\boldsymbol{p}_B+\boldsymbol{p}_C-\boldsymbol{p}_A){\cal{M}}^{M_{J_A}M_{J_B}M_{J_C}} = \nonumber\\
&&- \gamma \sqrt{8E_AE_BE_C} \sum_{M_{\rho_A}} \sum_{M_{L_A}} \sum_{M_{\rho_B}} \sum_{M_{L_B}} \sum_{M_{S_1}, M_{S_3}, M_{S_4}, m}\nonumber\\
&& \times  \langle j_A M_{j_A}S_3M_{S_3}|J_AM_{J_A}\rangle \langle L_{\rho_A} M_{L_{\rho_A}}L_{\lambda_A}M_{L_{\lambda_A}}|L_AM_{L_A} \rangle \nonumber\\ && \times \langle L_A M_{L_A}S_{12}M_{S_{12}}|j_AM_{j_A}\rangle \langle S_1M_{S_1}S_2M_{S_2}|S_{12}M_{S_{12}} \rangle \nonumber\\
&& \times \langle j_B M_{j_B}S_3M_{S_3}|J_BM_{J_B}\rangle \langle L_{\rho_B} M_{L_{\rho_B}}L_{\lambda_B}M_{L_{\lambda_B}}|L_BM_{L_B}\rangle \nonumber\\ && \times \langle L_B M_{L_B}S_{14}M_{S_{14}}|j_BM_{j_B}\rangle \langle S_1M_{S_1}S_4M_{S_4}|S_{14}M_{S_{14}}\rangle\nonumber\\
&& \times \langle 1m 1-m|00\rangle \langle S_4M_{S_4}S_5M_{S_5}|1-m \rangle \nonumber\\
&& \times \langle L_C M_{L_C}S_CM_{S_C}|J_CM_{J_C}\rangle \langle S_2M_{S_2}S_5M_{S_5}|S_CM_{S_C}\rangle \nonumber\\
&& \times \langle \phi_B^{143} \phi_C^{25}|\phi_A^{123}\phi_0^{45}\rangle I^{M_{L_A}m}_{M_{L_B}M_{L_C}}(\boldsymbol{p}),
\end{eqnarray}
where $\langle \phi_B^{143} \phi_C^{25}|\phi_A^{123}\phi_0^{45}\rangle$ and $I^{M_{L_A}m}_{M_{L_B}M_{L_C}}(\boldsymbol{p})$ are the overlaps of the flavor part and space part, respectively. Here, the overlap of space part $I^{M_{L_A}m}_{M_{L_B}M_{L_C}}(\boldsymbol{p})$ can be expressed as
\begin{eqnarray}
I^{M_{L_A}m}_{M_{L_B}M_{L_C}}(\boldsymbol{p}) & = & \int d^3\boldsymbol{p}_1d^3\boldsymbol{p}_2d^3\boldsymbol{p}_3d^3\boldsymbol{p}_4d^3\boldsymbol{p}_5  \nonumber\\ && \times \delta^3(\boldsymbol{p}_1+\boldsymbol{p}_2+\boldsymbol{p}_3-\boldsymbol{P}_A)\delta^3(\boldsymbol{p}_4+\boldsymbol{p}_5)\nonumber\\ && \times \delta^3(\boldsymbol{p}_1+\boldsymbol{p}_4+\boldsymbol{p}_3-\boldsymbol{P}_B)\delta^3(\boldsymbol{p}_2+\boldsymbol{p}_5-\boldsymbol{P}_C) \nonumber\\
&& \times \psi^*_B(\boldsymbol{p}_1,\boldsymbol{p}_4,\boldsymbol{p}_3) \psi^*_C(\boldsymbol{p}_2,\boldsymbol{p}_5)\nonumber\\
&& \times\psi_A(\boldsymbol{p}_1,\boldsymbol{p}_2,\boldsymbol{p}_3){\cal{Y}}^m_1\left(\frac{\boldsymbol{p}_4-\boldsymbol{p}_5}{2}\right
).
\end{eqnarray}
It should be mentioned that the overlap of space part not only relies on the magnetic quantum numbers, but also depends on the radial and orbital quantum numbers which are usually omitted for simplicity.

Then, the decay width
$\Gamma(A\rightarrow BC)$ can be calculated straightforwardly
\begin{eqnarray}
\Gamma= \pi^2\frac{p}{M^2_A}\frac{s}{2J_A+1}\sum_{M_{J_A},M_{J_B},M_{J_C}}|{\cal{M}}^{M_{J_A}M_{J_B}M_{J_C}}|^2,
\end{eqnarray}
where $p=|\boldsymbol{p}|=\frac{\sqrt{[M^2_A-(M_B+M_C)^2][M^2_A-(M_B-M_C)^2]}}{2M_A}$ is the momentum of the final states. The statistical factor $s=1/(1+\delta_{BC})$  always equals to one for $B$ and $C$ cannot be identical particles in present calculations.

The explicit notations of initial excited singly heavy baryons together with their predicted masses~\cite{Ebert:2011kk} are presented in Tab.~\ref{tab1}. For the initial higher excited states, the predicted masses are adopted to calculate their strong decay behaviors. For the final ground states, their masses are taken from the Review of Particle Physics~\cite{Tanabashi:2018oca}. The harmonic oscillator wave functions are employed to estimate the overlap of space part. For the harmonic oscillator parameters of different mesons, the effective values are used as in Ref.~\cite{Godfrey:2015dva}. For the baryon parameters, we use $\alpha_\rho=400~\rm{MeV}$ and
\begin{eqnarray}
\alpha_\lambda=\Bigg(\frac{3m_Q}{2m_q+m_Q} \Bigg)^{1/4} \alpha_\rho,
\end{eqnarray}
where the $m_Q$ and $m_q$ are the heavy and light quark masses, respectively~\cite{Liang:2019aag,Lu:2018utx,Zhong:2007gp}. The $m_{u/d}=220~\rm{MeV}$, $m_s=419~\rm{MeV}$, $m_c=1628~\rm{MeV}$, and $m_b=4977~\rm{MeV}$ are introduced to take into account the mass differences of heavy and light quarks~\cite{Godfrey:1985xj,Capstick:1986bm,Godfrey:2015dva}.
The overall parameter $\gamma$, can be determined by the well established $\Sigma_c(2520)^{++} \to \Lambda_c \pi^+$
process. The $\gamma=9.83$ can reproduce the width $\Gamma [\Sigma_c(2520)^{++} \to \Lambda_c \pi^+]=14.78~\rm{MeV}$~\cite{Tanabashi:2018oca,Lu:2018utx}, and also describe the decay widths of ground states $\Sigma_b$ and $\Sigma_b^*$ to $\Lambda_b \pi$ well~\cite{Liang:2019aag}.

\begin{table*}[!htbp]
\begin{center}
\caption{ \label{tab1} Notations, quantum numbers, and the predicted masses of the relevant excited singly heavy baryons. The $n_\rho$ and $L_\rho$ denote the radial and orbital angular momenta between the two light quarks, respectively, while $n_\lambda$ and $L_\lambda$ correspond to the radial and angular momenta between the light quark system and heavy quark, respectively. $L$ is the total orbital angular momentum, $S_\rho$ is the total spin of the light quarks, $j$ is total angular momentum of light quark system, $J$ is the total angular momentum, and $P$ stands for the parity. The predicted masses of these higher excited singly heavy baryons are taken from the relativistic quark model~\cite{Ebert:2011kk}. The units are in MeV.}
\renewcommand{\arraystretch}{1.5}
\begin{tabular*}{18cm}{@{\extracolsep{\fill}}*{11}{p{1.5cm}<{\centering}}}
\hline\hline
 State                          & $n_\rho$ & $L_\rho$     &  $n_\lambda$      &  $L_\lambda$  &  $L$  & $S_\rho$ & $j$  & $J^P$  & Charmed & Bottom        \\\hline
 $\Lambda_Q(3S)$     & 0        & 0            &  2                &  0            &  0    & 0        & 0      & $\frac{1}{2}^+$  & 3130  & 6455 \\
 $\Sigma_Q(3S)$      & 0        & 0            &  2                &  0            &  0    & 1        & 1      & $\frac{1}{2}^+$ & 3271  & 6575\\
 $\Sigma_Q^*(3S)$    & 0        & 0            &  2                &  0            &  0    & 1        & 1      & $\frac{3}{2}^+$ & 3293  & 6583\\
 $\Lambda_{Q1}(\frac{1}{2}^-,2P)$     & 0        & 0            &  1                &  1            &  1    & 0        & 1      & $\frac{1}{2}^-$ & 2983  & 6326 \\
 $\Lambda_{Q1}(\frac{3}{2}^-,2P)$     & 0        & 0            &  1                &  1            &  1    & 0        & 1      & $\frac{3}{2}^-$ & 3005  & 6333 \\
 $\Sigma_{Q0}(\frac{1}{2}^-,2P)$     & 0        & 0            &  1                &  1            &  1    & 1        & 0      & $\frac{1}{2}^-$ & 3172  & 6440 \\
 $\Sigma_{Q1}(\frac{1}{2}^-,2P)$     & 0        & 0            &  1                &  1            &  1    & 1        & 1      & $\frac{1}{2}^-$ & 3125  & 6430 \\
 $\Sigma_{Q1}(\frac{3}{2}^-,2P)$     & 0        & 0            &  1                &  1            &  1    & 1        & 1      & $\frac{3}{2}^-$ & 3172  & 6430 \\
 $\Sigma_{Q2}(\frac{3}{2}^-,2P)$     & 0        & 0            &  1                &  1            &  1    & 1        & 2      & $\frac{3}{2}^-$ & 3151  & 6423 \\
 $\Sigma_{Q2}(\frac{5}{2}^-,2P)$     & 0        & 0            &  1                &  1            &  1    & 1        & 2      & $\frac{5}{2}^-$ & 3161  & 6421 \\
 $\Lambda_{Q2}(\frac{3}{2}^+,2D)$     & 0        & 0            &  1                &  2            &  2    & 0        & 2      & $\frac{3}{2}^+$ & 3189  & 6526 \\
 $\Lambda_{Q2}(\frac{5}{2}^+,2D)$     & 0        & 0            &  1                &  2            &  2    & 0        & 2      & $\frac{5}{2}^+$ & 3209  & 6531 \\
 $\Sigma_{Q1}(\frac{1}{2}^+,2D)$     & 0        & 0            &  1                &  2            &  2    & 1        & 1      & $\frac{1}{2}^+$ & 3370  & 6636 \\
 $\Sigma_{Q1}(\frac{3}{2}^+,2D)$     & 0        & 0            &  1                &  2            &  2    & 1        & 1      & $\frac{3}{2}^+$ & 3366  & 6647 \\
 $\Sigma_{Q2}(\frac{3}{2}^+,2D)$     & 0        & 0            &  1                &  2            &  2    & 1        & 2      & $\frac{3}{2}^+$ & 3364  & 6612 \\
 $\Sigma_{Q2}(\frac{5}{2}^+,2D)$     & 0        & 0            &  1                &  2            &  2    & 1        & 2      & $\frac{5}{2}^+$ & 3365  & 6612 \\
 $\Sigma_{Q3}(\frac{5}{2}^+,2D)$     & 0        & 0            &  1                &  2            &  2    & 1        & 3      & $\frac{5}{2}^+$ & 3349  & 6598 \\
 $\Sigma_{Q3}(\frac{7}{2}^+,2D)$     & 0        & 0            &  1                &  2            &  2    & 1        & 3      & $\frac{7}{2}^+$ & 3342  & 6590 \\
 $\Lambda_{Q3}(\frac{5}{2}^-,1F)$     & 0        & 0            &  0                &  3            &  3    & 0        & 3      & $\frac{5}{2}^-$ & 3097  & 6408 \\
 $\Lambda_{Q3}(\frac{7}{2}^-,1F)$     & 0        & 0            &  0                &  3            &  3    & 0        & 3      & $\frac{7}{2}^-$ & 3078  & 6411 \\
 $\Sigma_{Q2}(\frac{3}{2}^-,1F)$     & 0        & 0            &  0                &  3            &  3    & 1        & 2      & $\frac{3}{2}^-$ & 3288  & 6550 \\
 $\Sigma_{Q2}(\frac{5}{2}^-,1F)$     & 0        & 0            &  0                &  3            &  3    & 1        & 2      & $\frac{5}{2}^-$ & 3283  & 6564 \\
 $\Sigma_{Q3}(\frac{5}{2}^-,1F)$     & 0        & 0            &  0                &  3            &  3    & 1        & 3      & $\frac{5}{2}^-$ & 3254  & 6501 \\
 $\Sigma_{Q3}(\frac{7}{2}^-,1F)$     & 0        & 0            &  0                &  3            &  3    & 1        & 3      & $\frac{7}{2}^-$ & 3253  & 6500 \\
 $\Sigma_{Q4}(\frac{7}{2}^-,1F)$     & 0        & 0            &  0                &  3            &  3    & 1        & 4      & $\frac{7}{2}^-$ & 3227  & 6472 \\
 $\Sigma_{Q4}(\frac{9}{2}^-,1F)$     & 0        & 0            &  0                &  3            &  3    & 1        & 4      & $\frac{9}{2}^-$ & 3209  & 6459 \\
 \hline\hline
\end{tabular*}
\end{center}
\end{table*}

\section{Strong decays of the higher $\Lambda_Q$ states}{\label{decay1}}

\subsection{$\Lambda_Q(3S)$}

In the constituent quark model, there is only one $\lambda-$mode $\Lambda_c(3S)$ state with $J^P=1/2^+$. Its mass is estimated to be $\sim 3130$ MeV within the relativistic quark model, which lies above the $D^{(*)}N$ thresholds. The strong decay behaviors are performed in Tab.~\ref{lamc3s}. The $\Lambda_c(3S)$ state may be a relatively narrow state with a width of $\sim 69$ MeV. Its decays may be governed by the $D^*N$ channel. The branching ratio of the $D^*N$ channel is predicted to be
\begin{eqnarray}
Br(\Lambda_c(3S) \to D^*N)  = 95\%,
\end{eqnarray}
which is independent with the overall parameter $\gamma$.

From the heavy flavor symmetry, the mass and decay behaviors of the $\Lambda_b(3S)$ state should be similar to that of the $\Lambda_c(3S)$. From Tab.~\ref{tab1}, the predicted mass gaps are
\begin{eqnarray}
m[\Lambda_c(3S)]- m[\Lambda_c]  = 844~\rm{MeV}
\end{eqnarray}
and
\begin{eqnarray}
m[\Lambda_b(3S)]- m[\Lambda_b]  = 835~\rm{MeV},
\end{eqnarray}
which preserve the heavy flavor symmetry well. From Tab.~\ref{lamb3s}, it can be seen that total decay width of $\Lambda_b(3S)$ is about 52 MeV, and the main decay mode is $B^*N$ channel. The branching ratio of $B^*N$ channel is predicted to be
\begin{eqnarray}
Br(\Lambda_b(3S) \to B^*N)  = 89\%.
\end{eqnarray}
It should be mentioned that the $\lambda-$mode $\Lambda_c(3S)$ and $\Lambda_b(3S)$ states mainly decay into the $D^*N$ and $B^*N$ channels, respectively, which indicate that these heavy meson plus light baryon final states should be ideal channels to to search for these $\Lambda_Q(3S)$ states in future experiments.

\begin{table}
\begin{center}
\caption{ \label{lamc3s} Decay widths of the $\Lambda_c(3S)$ state in MeV.}
\renewcommand{\arraystretch}{1.5}
\begin{tabular*}{8.0cm}{@{\extracolsep{\fill}}*{2}{p{3.5cm}<{\centering}}}
\hline\hline
   Mode                        & $\Lambda_c(3S)$     \\
  $\Sigma_c^{++} \pi^-$        & 0.11                  \\
  $\Sigma_c^{+} \pi^0$        & 0.11                  \\
  $\Sigma_c^{0} \pi^+$        & 0.11                \\
  $\Sigma_c^{*++} \pi^-$        & 0.27                    \\
  $\Sigma_c^{*+} \pi^0$        & 0.27                       \\
  $\Sigma_c^{*0} \pi^+$        & 0.27                      \\
  $\Lambda_c \omega$        & 0.32                          \\
  $\Xi_c^{\prime+} K^0$     & 0.08                              \\
  $\Xi_c^{\prime0} K^+$     & 0.08                                   \\
  $D^0p$                       & 0.04                             \\
  $D^+n$                       & 0.10                            \\
  $D^{*0}p$                       & 32.00                      \\
  $D^{*+}n$                       & 33.43                          \\
  $D_s \Lambda$                &  1.83                              \\
  Total width                  & 69.03                       \\
\hline\hline
\end{tabular*}
\end{center}
\end{table}

\begin{table}
\begin{center}
\caption{ \label{lamb3s} Decay widths of the $\Lambda_b(3S)$ state in MeV.}
\renewcommand{\arraystretch}{1.5}
\begin{tabular*}{8.0cm}{@{\extracolsep{\fill}}*{2}{p{3.5cm}<{\centering}}}
\hline\hline
   Mode                        & $\Lambda_b(3S)$     \\
  $\Sigma_b^{+} \pi^-$        & 0.19                  \\
  $\Sigma_b^{0} \pi^0$        & 0.19                  \\
  $\Sigma_b^{-} \pi^+$        & 0.19                \\
  $\Sigma_b^{*+} \pi^-$        & 0.40                    \\
  $\Sigma_b^{*0} \pi^0$        & 0.40                      \\
  $\Sigma_b^{*-} \pi^+$        & 0.41                      \\
  $\Lambda_b \omega$        & 0.47                          \\
  $\Xi_b^{\prime0} K^0$     & 0.02                              \\
  $\Xi_b^{\prime-} K^+$     & 0.03                                   \\
  $\Xi_b^{\prime*0} K^0$     & 0.01                              \\
  $\Xi_b^{\prime*-} K^+$     & 0.01                                   \\
  $B^-p$                       & 1.51                             \\
  $B^0n$                       & 1.61                            \\
  $B^{*-}p$                       & 22.81                      \\
  $B^{*0}n$                       & 23.39                          \\
  Total width                  & 51.65                       \\
\hline\hline
\end{tabular*}
\end{center}
\end{table}

\subsection{$\Lambda_Q(2P)$}

In the conventional quark model, there are two $\lambda-$type $\Lambda_c(2P)$ states with $J^P=1/2^-$ and $J^P=3/2^-$. The predicted masses in the relativistic quark model are around 2983 and 3005 MeV, respectively, and their strong decay widths are listed in Tab.~\ref{lamc2p}. It can be seen that the dominated decay channels of these two states are $D^*N$ and $DN$, while other decay modes can be neglected.

As mentioned in the Introduction, the $\Lambda_c(2940)$ is a good candidate of the $\Lambda_c(2P)$ states in the consideration of the mass uncertainties of quark model. The mass of $\Lambda_c(2940)$ lies below the $D^*N$ threshold, which leads to a narrow total decay width. In previous work~\cite{Lu:2018utx}, the analysis of the strong decays indicates that $\Lambda_c(2940)$ can be well explained as the $J^P=3/2^-$ $\Lambda_c(2P)$ state in the $N=3$ shell within the same model and parameters adopted in present work. The detailed discussions of these two $\lambda-$type $\Lambda_c(2P)$ states can be found in Ref.~\cite{Lu:2018utx}.

\begin{figure}[htb]
\includegraphics[scale=0.75]{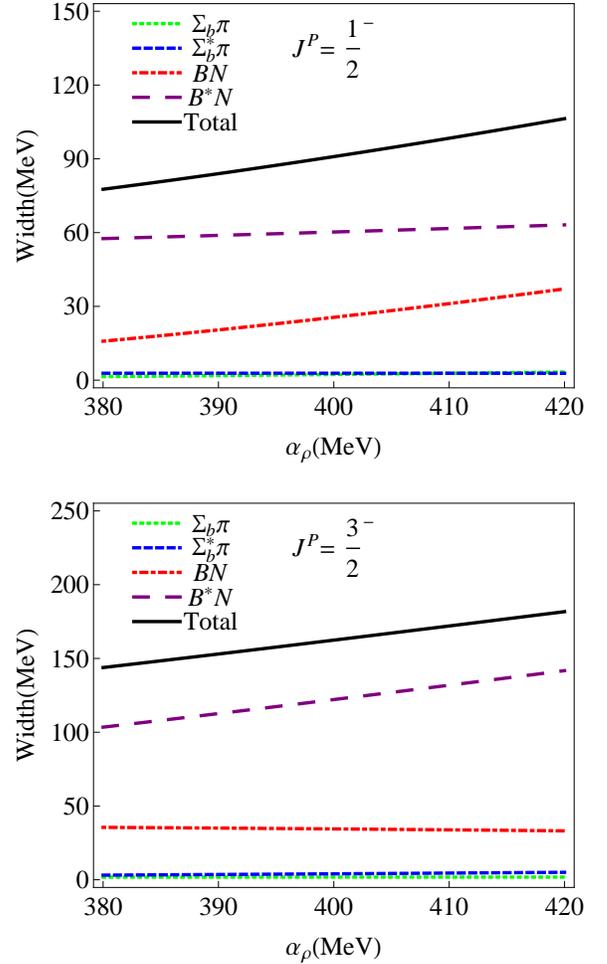}
\vspace{0.0cm} \caption{The decay widths of the $\Lambda_{b1}(\frac{1}{2}^-,2P)$ and $\Lambda_{b1}(\frac{3}{2}^-,2P)$ states as functions of the harmonic oscillator parameter $\alpha_\rho$.}
\label{lambdabp}
\end{figure}

\begin{figure}[htb]
\includegraphics[scale=0.75]{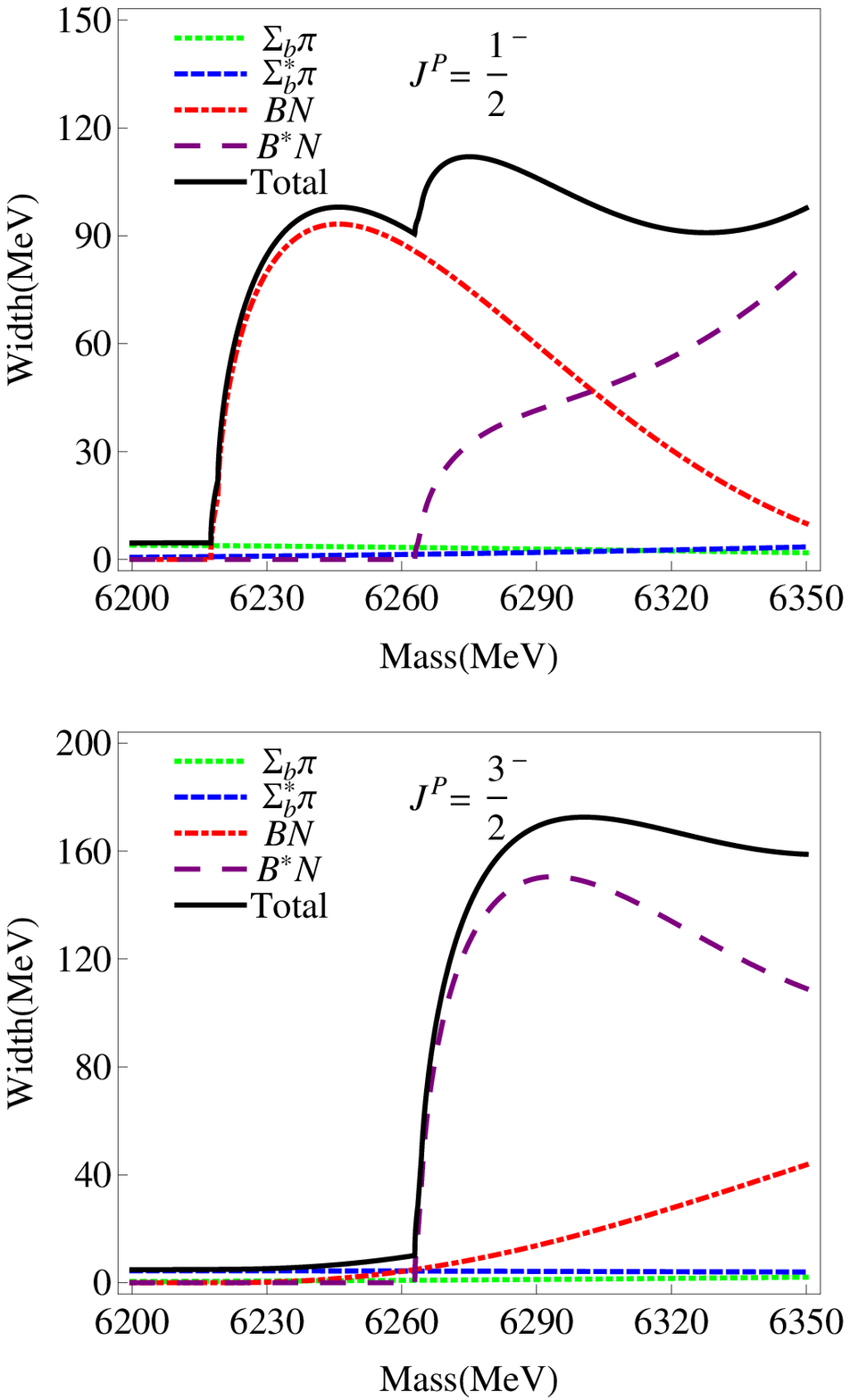}
\vspace{0.0cm} \caption{The decay widths of the $\Lambda_{b1}(\frac{1}{2}^-,2P)$ and $\Lambda_{b1}(\frac{3}{2}^-,2P)$ states as functions of the their masses.}
\label{lambdab}
\end{figure}

The strong decay behaviors of $\Lambda_b(2P)$ states with predicted masses are shown in Tab.~\ref{lamb2p}. The total decay widths are 91 and 162 MeV for $J^P=1/2^-$ and $J^P=3/2^-$ states, respectively. The main decay channels are $BN$ and $B^*N$. The variations of strong decays versus the harmonic oscillator parameter $\alpha_\rho$ are presented in Fig.~\ref{lambdabp}, which shows relatively stable predictions. It should be mentioned that if the $\Lambda_c(2940)$ corresponds to the $J^P=3/2^-$ $\Lambda_c(2P)$ state indeed, the masses of the $\Lambda_b(2P)$ may lie below the $B^*N$ threshold as well considering the heavy flavor symmetry. The variations of strong decays versus the initial $\Lambda_b(2P)$ masses are shown in Fig.~\ref{lambdab}. When the initial masses decrease below the $B^*N$ threshold, the total decay width of $J^P=1/2^-$ states approximately remain because of the large $BN$ channel. However, for the $J^P=3/2^-$ state, the total decay width drops into several MeV if its mass is lower than the $B^*N$ threshold. This specific feature of $\Lambda_b(2P)$ states may provide helpful information for future experimental searches.

\begin{table}
\begin{center}
\caption{ \label{lamc2p} Decay widths of the $\Lambda_c(2P)$ states in MeV.}
\renewcommand{\arraystretch}{1.5}
\begin{tabular*}{8.5cm}{@{\extracolsep{\fill}}*{3}{p{2.5cm}<{\centering}}}
\hline\hline
   Mode                        & $\Lambda_{c1}(\frac{1}{2}^-,2P)$      & $\Lambda_{c1}(\frac{3}{2}^-,2P)$     \\
  $\Sigma_c^{++} \pi^-$        & 0.49                                  & 0.52             \\
  $\Sigma_c^{+} \pi^0$        & 0.48                                   & 0.53             \\
  $\Sigma_c^{0} \pi^+$        & 0.49                                   & 0.52             \\
  $\Sigma_c^{*++} \pi^-$        & 0.50                                 & 0.97             \\
  $\Sigma_c^{*+} \pi^0$        & 0.51                                  & 0.97             \\
  $\Sigma_c^{*0} \pi^+$        & 0.50                                  & 0.97             \\
  $D^0p$                       & 1.19                                  & 19.54          \\
  $D^+n$                       & 1.68                                  & 18.80              \\
  $D^{*0}p$                       & 17.31                              & 55.25          \\
  $D^{*+}n$                       & 16.67                              & 56.59              \\
  Total width                  & 39.84                                 & 154.67       \\
\hline\hline
\end{tabular*}
\end{center}
\end{table}

\begin{table}
\begin{center}
\caption{ \label{lamb2p} Decay widths of the $\Lambda_b(2P)$ states in MeV.}
\renewcommand{\arraystretch}{1.5}
\begin{tabular*}{8.5cm}{@{\extracolsep{\fill}}*{3}{p{2.5cm}<{\centering}}}
\hline\hline
   Mode                        & $\Lambda_{b1}(\frac{1}{2}^-,2P)$      & $\Lambda_{b1}(\frac{3}{2}^-,2P)$     \\
  $\Sigma_b^{+} \pi^-$        & 0.76                                  & 0.60             \\
  $\Sigma_b^{0} \pi^0$        & 0.75                                   & 0.61             \\
  $\Sigma_b^{-} \pi^+$        & 0.78                                   & 0.58             \\
  $\Sigma_b^{*+} \pi^-$        & 0.94                                 & 1.34            \\
  $\Sigma_b^{*0} \pi^0$        & 0.95                                  & 1.34             \\
  $\Sigma_b^{*-} \pi^+$        & 0.92                                  & 1.35             \\
  $B^-p$                       & 12.47                                  & 17.42          \\
  $B^0n$                       & 13.07                                  & 17.04              \\
  $B^{*-}p$                       & 30.33                              & 60.76          \\
  $B^{*0}n$                       & 29.91                              & 61.37              \\
  Total width                  & 90.90                                 & 162.42       \\
\hline\hline
\end{tabular*}
\end{center}
\end{table}

\subsection{$\Lambda_Q(2D)$}

The predicted masses of the two $\Lambda_c(2D)$ states with $J^P=3/2^+$ and $J^P=5/2^+$ are around 3189 and 3209 MeV, respectively. Until now, there is no experimental evidence of the $\Lambda_c$ states above 3000 MeV. We employe the predicted masses to calculate their strong decay widths, and the results are performed in Tab.~\ref{lamc2d}. The total decay widths of these two states are about 57 and 73 MeV, respectively. It is shown that the dominated decay channel is $D^*N$ for the $J^P=3/2^+$ state, while the main decay modes are both $DN$ and $D^*N$ final states for the $J^P=5/2^+$ states. The partial decay width ratios are predicted to be
\begin{small}
\begin{eqnarray}
\Gamma[\Lambda_{c2}\left(\frac{3}{2}^+,2D\right) \to DN]:\Gamma[\Lambda_{c2}\left(\frac{3}{2}^+,2D\right) \to D^*N]  = 9\times 10^{-3}
\end{eqnarray}
\end{small}
and
\begin{small}
\begin{eqnarray}
\Gamma[\Lambda_{c2}\left(\frac{5}{2}^+,2D\right) \to DN]:\Gamma[\Lambda_{c2}\left(\frac{5}{2}^+,2D\right) \to D^*N]  = 0.63.
\end{eqnarray}
\end{small}
These different partial decay ratios can help us to distinguish the two $\Lambda_c(2D)$ states.

In the bottom sector, the similar situation happens. The masses and total decay widths of the two $\Lambda_b(2D)$ states are almost same. The dominant decay channel is $B^*N$ for the $J^P=3/2^+$ state, while the main decay modes are $BN$ and $B^*N$ final states for the $J^P=5/2^+$ states. To distinguish these two states, the partial decay ratios are needed. From Tab.~\ref{lamb2d}, it is shown that
\begin{small}
\begin{eqnarray}
\Gamma[\Lambda_{b2}\left(\frac{3}{2}^+,2D\right) \to BN]:\Gamma[\Lambda_{b2}\left(\frac{3}{2}^+,2D\right) \to B^*N]  = 4\times 10^{-3}
\end{eqnarray}
\end{small}
and
\begin{small}
\begin{eqnarray}
\Gamma[\Lambda_{b2}\left(\frac{5}{2}^+,2D\right) \to BN]:\Gamma[\Lambda_{b2}\left(\frac{5}{2}^+,2D\right) \to B^*N]  = 0.70,
\end{eqnarray}
\end{small}
which in independent with the quark pair creation strengthen. To sum up, the $\lambda-$mode $\Lambda_Q(2D)$ states mainly decay into heavy meson plus light baryon final states, which can be tested by future experiments.

\begin{table}
\begin{center}
\caption{ \label{lamc2d} Decay widths of the $\Lambda_c(2D)$ states in MeV.}
\renewcommand{\arraystretch}{1.5}
\begin{tabular*}{8.5cm}{@{\extracolsep{\fill}}*{3}{p{2.5cm}<{\centering}}}
\hline\hline
   Mode                        & $\Lambda_{c2}(\frac{3}{2}^+,2D)$      & $\Lambda_{c2}(\frac{5}{2}^+,2D)$     \\
  $\Sigma_c^{++} \pi^-$        & 0.07                                  & 0.08             \\
  $\Sigma_c^{+} \pi^0$        & 0.07                                   & 0.08             \\
  $\Sigma_c^{0} \pi^+$        & 0.07                                   & 0.08             \\
  $\Sigma_c^{*++} \pi^-$        & 0.10                                 & 0.14             \\
  $\Sigma_c^{*+} \pi^0$        & 0.10                                  & 0.14             \\
  $\Sigma_c^{*0} \pi^+$        & 0.10                                  & 0.14             \\
  $\Lambda_c \omega$        & 0.17                                     & 0.19             \\
  $\Xi_c^{\prime+} K^0$     & 0.09                                     & $2\times 10^{-3}$                 \\
  $\Xi_c^{\prime0} K^+$     & 0.10                                     & $2\times 10^{-3}$                 \\
  $\Xi_c^{\prime*+} K^0$     & 0.01                                    & 0.06                 \\
  $\Xi_c^{\prime*0} K^+$     & 0.01                                    & 0.06                 \\
  $D^0p$                       & 0.27                                  & 13.85          \\
  $D^+n$                       & 0.18                                  & 13.69              \\
  $D^{*0}p$                       & 26.69                              & 22.00          \\
  $D^{*+}n$                       & 26.02                              & 22.03              \\
  $D_s \Lambda$                &  3.29                                 & 0.46                   \\
  Total width                  & 57.32                                 & 73.00       \\
\hline\hline
\end{tabular*}
\end{center}
\end{table}

\begin{table}
\begin{center}
\caption{ \label{lamb2d} Decay widths of the $\Lambda_b(2D)$ states in MeV.}
\renewcommand{\arraystretch}{1.5}
\begin{tabular*}{8.5cm}{@{\extracolsep{\fill}}*{3}{p{2.5cm}<{\centering}}}
\hline\hline
   Mode                        & $\Lambda_{b2}(\frac{3}{2}^+,2D)$      & $\Lambda_{b2}(\frac{5}{2}^+,2D)$     \\
  $\Sigma_b^{+} \pi^-$        & 0.10                                  & 0.10             \\
  $\Sigma_b^{0} \pi^0$        & 0.10                                   & 0.10             \\
  $\Sigma_b^{-} \pi^+$        & 0.10                                   & 0.10             \\
  $\Sigma_b^{*+} \pi^-$        & 0.18                                 & 0.20             \\
  $\Sigma_b^{*0} \pi^0$        & 0.18                                  & 0.20             \\
  $\Sigma_b^{*-} \pi^+$        & 0.18                                  & 0.20             \\
  $\Lambda_b \omega$        & 0.29                                     & 0.30             \\
  $\Xi_b^{\prime0} K^0$     & 0.09                                     & $8\times 10^{-4}$                 \\
  $\Xi_b^{\prime-} K^+$     & 0.09                                     & $9\times 10^{-4}$                 \\
  $\Xi_b^{\prime*0} K^0$     & 0.01                                    & 0.09                 \\
  $\Xi_b^{\prime*-} K^+$     & 0.01                                    & 0.09                 \\
  $B^-p$                       & 0.21                                  & 21.77          \\
  $B^0n$                       & 0.18                                  & 21.69              \\
  $B^{*-}p$                       & 50.61                              & 30.91          \\
  $B^{*0}n$                       & 50.35                              & 30.82              \\
  $B_s \Lambda$                &  4.40                                 & 0.09                   \\
  Total width                  & 107.06                                & 106.67       \\
\hline\hline
\end{tabular*}
\end{center}
\end{table}

\subsection{$\Lambda_Q(1F)$}

For the $\Lambda_c(1F)$ states, the predicted masses are around 3097 and 3078 MeV for the $J^P=5/2^-$ and $J^P=7/2^-$ states, respectively. The calculated strong decay widths are listed in Tab.~\ref{lamc1f}. It is shown that the $DN$ and $D^*N$ are the dominated decay modes for these $F-$wave states, while the contribution of heavy baryon plus light meson final states can be neglected. Since the heavy flavor symmetry, the $\Lambda_b(1F)$ states have the similar properties to the charm sector. We perform their strong decay behaviors in Tab.~\ref{lamb1f} for reference.

\begin{figure}[htb]
\includegraphics[scale=0.75]{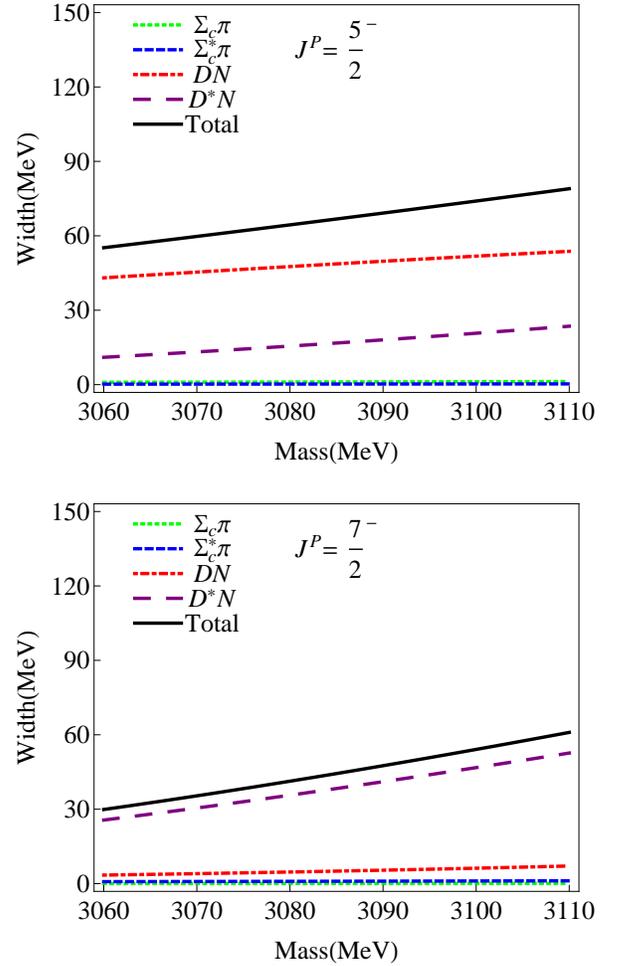}
\vspace{0.0cm} \caption{The decay widths of the $\Lambda_{c3}(\frac{5}{2}^-,1F)$ and $\Lambda_{c3}(\frac{7}{2}^-,1F)$ states as functions of the their masses. The partial decay widths of $\Lambda_c \omega$, $\Xi_c^{\prime+} K^0$, $\Xi_c^{\prime0} K^+$, and $D_s \Lambda$ channels are relatively small, which are
not presented here.}
\label{lambdac1f}
\end{figure}

\begin{figure}[htb]
\includegraphics[scale=0.75]{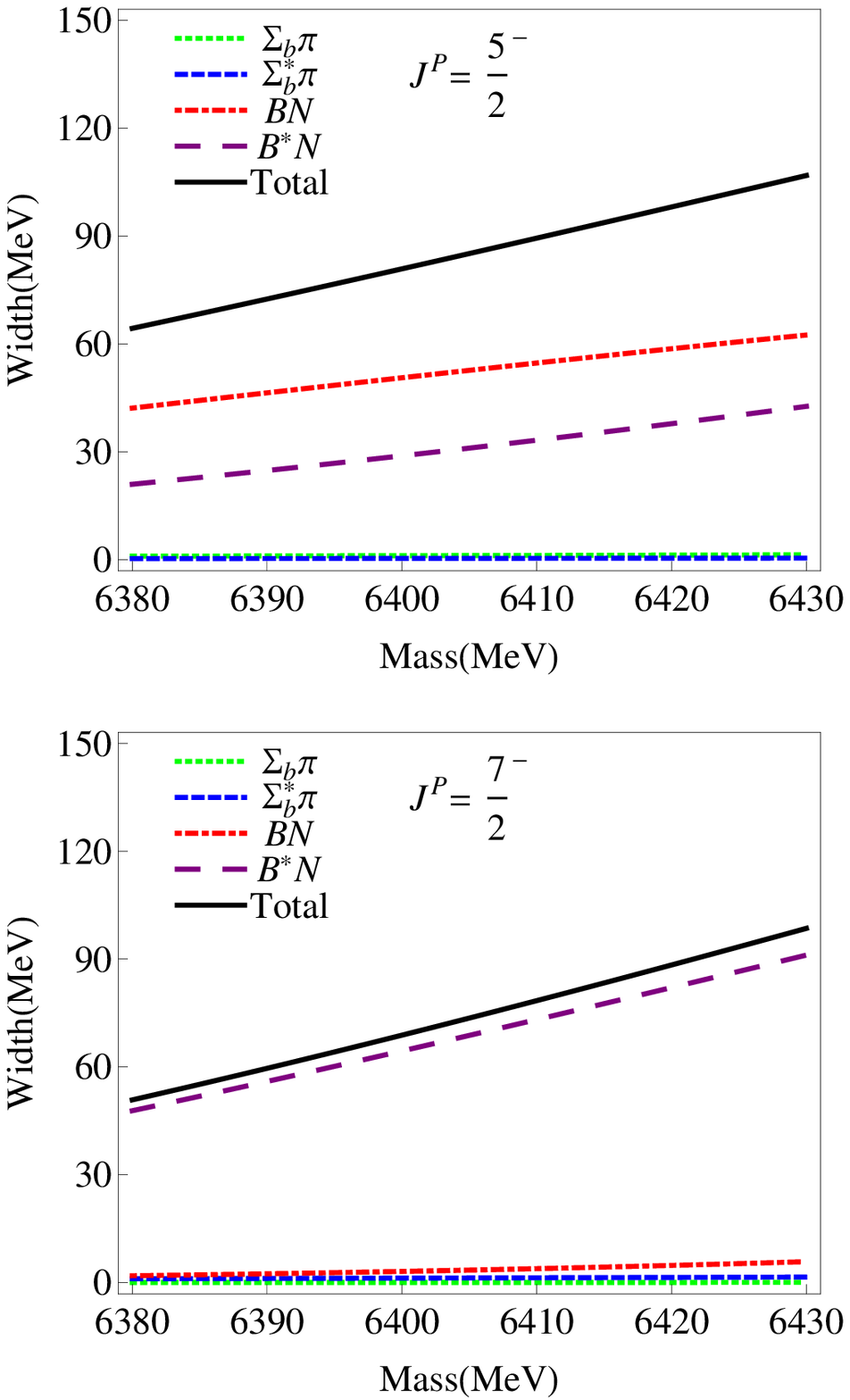}
\vspace{0.0cm} \caption{The decay widths of the $\Lambda_{b3}(\frac{5}{2}^-,1F)$ and $\Lambda_{b3}(\frac{5}{2}^-,1F)$ states as functions of the their masses. The partial decay width of $\Lambda_b \omega$ channel is relatively small, which is not presented here.}
\label{lambdab1f}
\end{figure}

\begin{figure}[htb]
\includegraphics[scale=0.75]{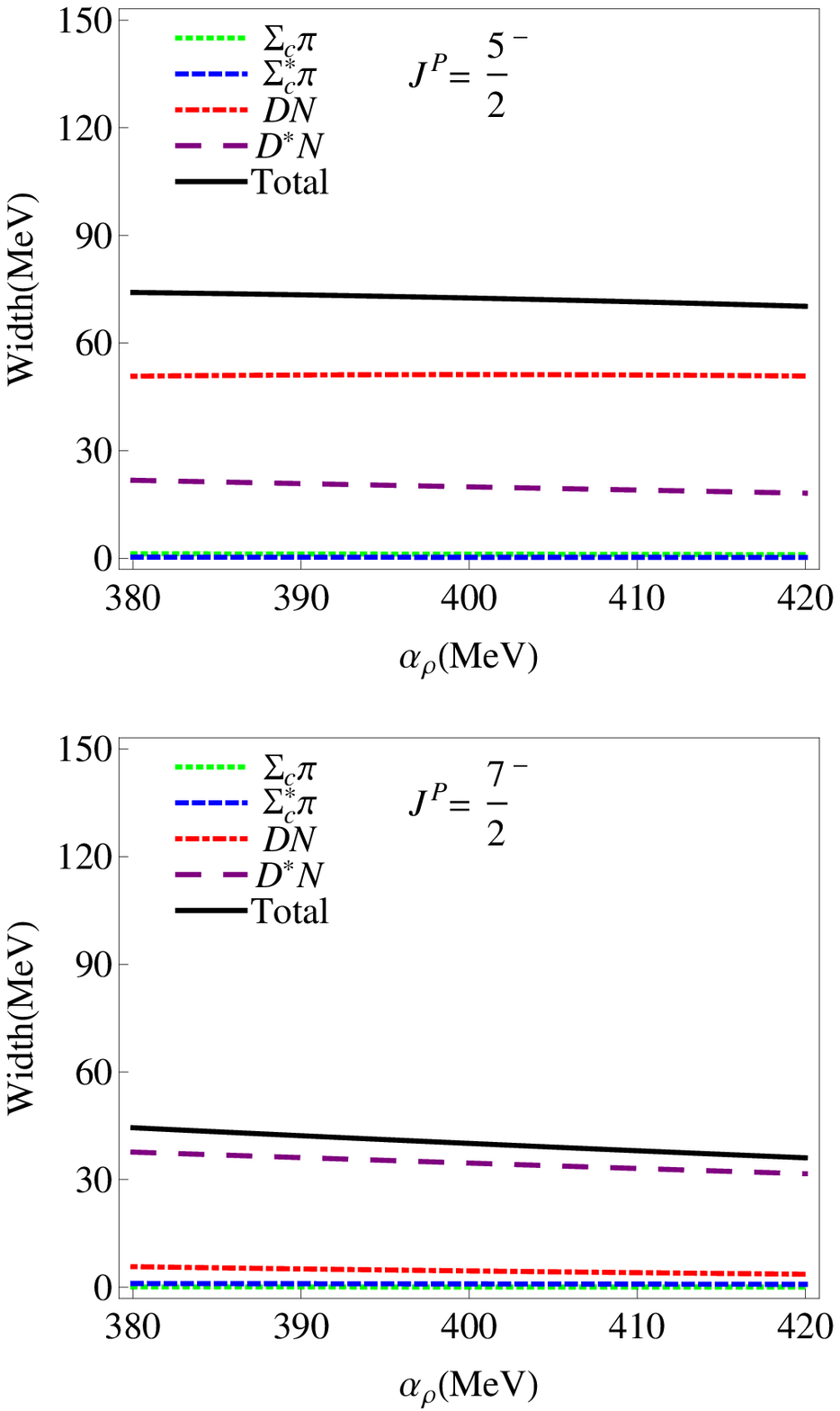}
\vspace{0.0cm} \caption{The decay widths of the $\Lambda_{c3}(\frac{5}{2}^-,1F)$ and $\Lambda_{c3}(\frac{7}{2}^-,1F)$ states as functions of the harmonic oscillator parameter $\alpha_\rho$. The partial decay widths of $\Lambda_c \omega$, $\Xi_c^{\prime+} K^0$, $\Xi_c^{\prime0} K^+$, and $D_s \Lambda$ channels are relatively small, which are
not presented here.}
\label{lambdac1fp}
\end{figure}

\begin{figure}[htb]
\includegraphics[scale=0.75]{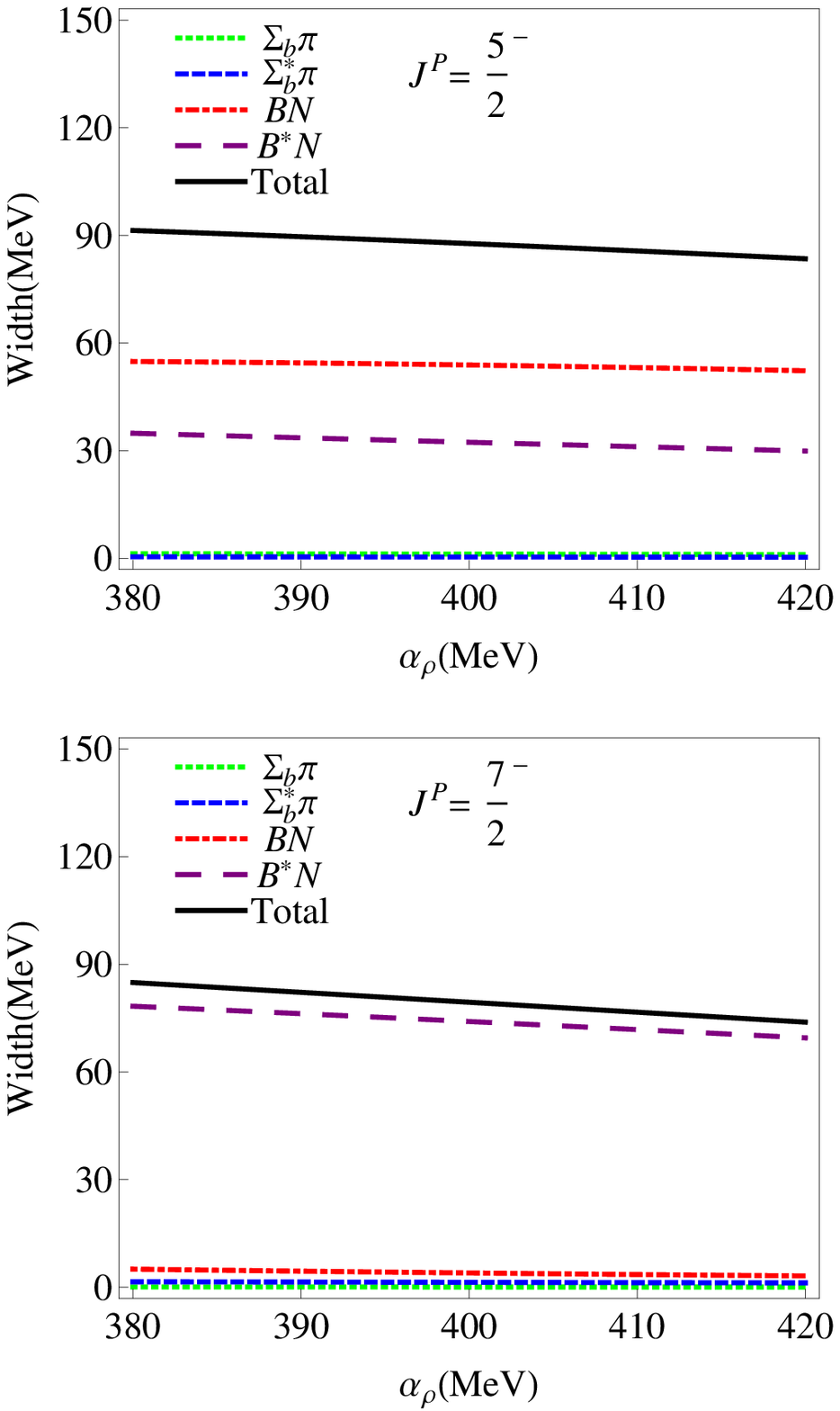}
\vspace{0.0cm} \caption{The decay widths of the $\Lambda_{b3}(\frac{5}{2}^-,1F)$ and $\Lambda_{b3}(\frac{5}{2}^-,1F)$ states as functions of the the harmonic oscillator parameter $\alpha_\rho$. The partial decay width of $\Lambda_b \omega$ channel is relatively small, which is not presented here.}
\label{lambdab1fp}
\end{figure}

From Tab.~\ref{tab1}, it can be found that the predicted masses of $\Lambda_Q(1F)$ states lie about $70-100$ MeV above the $\Lambda_Q(2P)$ states. The total decay widths are about $40-90$ MeV, and for the $\Lambda_c(1F)$ states the main decay channels are $D^{(*)}N$, while for the $\Lambda_b(1F)$ states, the main decay modes are $B^{(*)}N$. The variations of strong decays versus the initial $\Lambda_Q(1F)$ masses are also presented in Figs.~\ref{lambdac1f} and~\ref{lambdab1f}. It is shown that the total decay widths increase smoothly with the initial masses. With the 50 MeV mass uncertainties, the total decay widths vary in certain ranges. For instance, with the 50 MeV variations of initial masses, the total decay widths of two $\Lambda_b(1F)$ states lie in the range $64\sim 105$ MeV and $51 \sim 99$  MeV, respectively. Compared with the predicted central values 88 MeV and 79 MeV from Tab. IX, the uncertainties of total decay widths are up to 27\% and 35 \% for the $\Lambda_{b3}(\frac{5}{2}^-,1F)$ and $\Lambda_{b3}(\frac{7}{2}^-,1F)$ states, respectively. Moreover, the uncertainties raising from the harmonic oscillator parameter $\alpha_\rho$ are calculated. From Figs.~\ref{lambdac1fp} and~\ref{lambdab1fp}, it can be seen that our results are rather stable when the parameter $\alpha_\rho$ varies in a reasonable range. Considering the $\Lambda_c(2940)$ observed in the $D^0p$ invariant mass by the LHCb Collaboration, we hope further experiments can be carried out to search for these $\Lambda_Q(1F)$ states in the $D^{(*)}N$ and $B^{(*)}N$ final states.

\begin{table}
\begin{center}
\caption{ \label{lamc1f} Decay widths of the $\Lambda_c(1F)$ states in MeV.}
\renewcommand{\arraystretch}{1.5}
\begin{tabular*}{8.5cm}{@{\extracolsep{\fill}}*{3}{p{2.5cm}<{\centering}}}
\hline\hline
   Mode                        & $\Lambda_{c3}(\frac{5}{2}^-,1F)$      & $\Lambda_{c3}(\frac{7}{2}^-,1F)$     \\
  $\Sigma_c^{++} \pi^-$        & 0.38                                  & 0.02             \\
  $\Sigma_c^{+} \pi^0$        & 0.39                                   & 0.02             \\
  $\Sigma_c^{0} \pi^+$        & 0.38                                   & 0.02             \\
  $\Sigma_c^{*++} \pi^-$        & 0.09                                 & 0.30             \\
  $\Sigma_c^{*+} \pi^0$        & 0.09                                  & 0.30             \\
  $\Sigma_c^{*0} \pi^+$        & 0.09                                  & 0.30             \\
  $\Lambda_c \omega$        & 0.01                                     & $3\times 10^{-4}$             \\
  $\Xi_c^{\prime+} K^0$     & $4\times 10^{-4}$                        & $6\times 10^{-12}$                 \\
  $\Xi_c^{\prime0} K^+$     & $6\times 10^{-4}$                        & $9\times 10^{-11}$                 \\
  $D^0p$                       & 25.86                                 & 2.36          \\
  $D^+n$                       & 25.34                                 & 2.18              \\
  $D^{*0}p$                       & 10.25                              & 17.88          \\
  $D^{*+}n$                       & 9.65                               & 16.71              \\
  $D_s \Lambda$                &  0.02                             & $\cdots$                   \\
  Total width                  & 72.54                                 & 40.08       \\
\hline\hline
\end{tabular*}
\end{center}
\end{table}

\begin{table}
\begin{center}
\caption{ \label{lamb1f} Decay widths of the $\Lambda_b(1F)$ states in MeV.}
\renewcommand{\arraystretch}{1.5}
\begin{tabular*}{8.5cm}{@{\extracolsep{\fill}}*{3}{p{2.5cm}<{\centering}}}
\hline\hline
   Mode                        & $\Lambda_{b3}(\frac{5}{2}^-,1F)$      & $\Lambda_{b3}(\frac{7}{2}^-,1F)$     \\
  $\Sigma_b^{+} \pi^-$        & 0.39                                  & 0.02             \\
  $\Sigma_b^{0} \pi^0$        & 0.39                                   & 0.02             \\
  $\Sigma_b^{-} \pi^+$        & 0.38                                   & 0.02             \\
  $\Sigma_b^{*+} \pi^-$        & 0.12                                 & 0.45             \\
  $\Sigma_b^{*0} \pi^0$        & 0.12                                  & 0.45             \\
  $\Sigma_b^{*-} \pi^+$        & 0.12                                  & 0.44             \\
  $\Lambda_b \omega$        & $2\times 10^{-4}$                       & $5\times 10^{-4}$             \\
  $B^-p$                       & 27.07                                 & 2.01          \\
  $B^0n$                       & 26.81                                 & 1.95              \\
  $B^{*-}p$                       & 16.28                              & 37.26          \\
  $B^{*0}n$                       & 16.04                               & 36.79              \\
  Total width                  & 87.71                                 & 79.42       \\
\hline\hline
\end{tabular*}
\end{center}
\end{table}

\section{Strong decays of the higher $\Sigma_Q$ states}{\label{decay2}}

\subsection{$\Sigma_Q(3S)$}

In the constituent quark model, there are two $\lambda-$mode $3S$ $\Sigma_c$ states, $\Sigma_c(3S)$ and $\Sigma_c^*(3S)$. From Tab.~\ref{tab1}, the predicted masses of $\Sigma_c(3S)$ and $\Sigma_c^*(3S)$ are around 3271 and 3293 MeV, respectively. With this predicted masses, their strong decays are presented in Tab.~\ref{sigc3s}. Our results show that the total decay widths of $\Sigma_c(3S)$ and $\Sigma_c^*(3S)$ states are 58 and 89 MeV, respectively. The main decay channels of these two $3S$ $\Sigma_c$ states are $D\Delta$ and $D^*\Delta$ other than $DN$ and $D^*N$. For the $\Sigma_c(3S)$ state, the branching ratios of $DN$, $D^*N$, $D\Delta$, and $D^*\Delta$ modes are predicted to be
\begin{small}
\begin{eqnarray}
Br(DN, D^*N, D\Delta, D^*\Delta)  = 0.67\%, 0.03\%, 27.29\%, 66.27\%.
\end{eqnarray}
\end{small}
For the $\Sigma_c^*(3S)$ state, the branching ratios of $DN$, $D^*N$, $D\Delta$, and $D^*\Delta$ modes are predicted to be
\begin{small}
\begin{eqnarray}
Br(DN, D^*N, D\Delta, D^*\Delta)  = 2.33\%, 0.11\%, 7.97\%, 85.92\%.
\end{eqnarray}
\end{small}
These branching ratios are independent on the overall $\gamma$ and can be employed to distinguish these two states.

The predicted masses of $\Sigma_b(3S)$ and $\Sigma_b^*(3S)$ are around 6575 and 6583 MeV, respectively. From Tab.~\ref{sigb3s}, it is shown that the total decay widths of $\Sigma_b(3S)$ and $\Sigma_b^*(3S)$ states are 87 and 104 MeV, respectively. The dominated decay channels of these two bottom baryons are $B\Delta$ and $B^*\Delta$. For the $\Sigma_b(3S)$ state, the branching ratios of $BN$, $B^*N$, $B\Delta$, and $B^*\Delta$ modes are
\begin{small}
\begin{eqnarray}
Br(BN, B^*N, B\Delta, B^*\Delta)  = 0.48\%, 1.53\%, 52.14\%, 42.84\%.
\end{eqnarray}
\end{small}
For the $\Sigma_b^*(3S)$ state, the branching ratios of $BN$, $B^*N$, $B\Delta$, and $B^*\Delta$ modes are predicted to be
\begin{small}
\begin{eqnarray}
Br(BN, B^*N, B\Delta, B^*\Delta)  = 1.89\%, 1.35\%, 28.23\%, 64.92\%.
\end{eqnarray}
\end{small}
These branching ratios suggest that the $B\Delta$ and $B^*\Delta$ final states are the ideal channels to hunt for the $\Sigma_b(3S)$ and $\Sigma_b^*(3S)$ states. Moreover, the heavy flavor symmetry is approximately preserved by comparing the decay behaviors between the charmed and bottom sectors.

\begin{table}
\begin{center}
\caption{ \label{sigc3s} Decay widths of the $\Sigma_c(3S)$ states in MeV.}
\renewcommand{\arraystretch}{1.5}
\begin{tabular*}{8.5cm}{@{\extracolsep{\fill}}*{3}{p{2.5cm}<{\centering}}}
\hline\hline
 Mode      & $\Sigma_c^*(3S)$  & $\Sigma_{c}(3S)$    \\
 $\Lambda_c \pi^+$          & $1\times10^{-4}$               &  $2\times10^{-3}$              \\
 $\Lambda_c \rho^+$         & 0.42                     &  0.40               \\
 $\Sigma_c^{++} \pi^0$      & 0.06                     &  0.01              \\
 $\Sigma_c^{+} \pi^+$       & 0.06                     &  0.01             \\
 $\Sigma_c^{*++} \pi^0$     & 0.05                     &  0.10            \\
 $\Sigma_c^{*+} \pi^+$      & 0.05                     &  0.11               \\
 $\Sigma_c^{++} \rho^0$      & 0.39                    & 0.08                \\
 $\Sigma_c^{+} \rho^+$       & 0.41                    & 0.08                 \\
 $\Sigma_c^{*+} \rho^+$    & $\cdots$                  &  $5\times10^{-4}$               \\
 $\Sigma_c^{++} \eta$       & 0.12                     &  0.03              \\
 $\Sigma_c^{*++} \eta$      & 0.06                     &  0.16              \\
 $\Sigma_c^{++} \omega$       & 0.31                   &  0.07                \\
 $\Xi_c^{+} K^+$            & 0.42                    &  0.42               \\
 $\Xi_c^{\prime +} K^+$     & 0.47                     &  0.13            \\
 $\Xi_c^{\prime *+} K^+$    & 0.16                     &   0.47               \\
 $D^+ p$                    & 0.39                    &  2.07              \\
 $D^{*+} p$                 & 0.02                    &  0.10              \\
 $D^0 \Delta^{++}$          & 11.68                    &  5.20              \\
 $D^+ \Delta^{+}$           & 4.12                    &  1.87              \\
 $D^{*0} \Delta^{++}$          & 29.64                  &   57.85             \\
 $D^{*+} \Delta^{+}$           & 8.73                  &   18.33                \\
 $D_s \Sigma^{+}$           & 0.33                    &   1.20               \\
 Total                      & 57.90                   & 88.66               \\
\hline\hline
\end{tabular*}
\end{center}
\end{table}

\begin{table}
\begin{center}
\caption{ \label{sigb3s} Decay widths of the $\Sigma_b(3S)$ states in MeV.}
\renewcommand{\arraystretch}{1.5}
\begin{tabular*}{8.5cm}{@{\extracolsep{\fill}}*{3}{p{2.5cm}<{\centering}}}
\hline\hline
 Mode     & $\Sigma_b^*(3S)$  & $\Sigma_b(3S)$ \\
 $\Lambda_b \pi^+$          & $4\times10^{-4}$ & $1\times10^{-3}$            \\
 $\Lambda_b \rho^+$         & 0.70            & 0.69             \\
 $\Sigma_b^{+} \pi^0$      & 0.13            & 0.03            \\
 $\Sigma_b^{0} \pi^+$       & 0.13            & 0.03           \\
 $\Sigma_b^{*+} \pi^0$     & 0.08            & 0.18            \\
 $\Sigma_b^{*0} \pi^+$      & 0.08            & 0.18            \\
 $\Sigma_b^{+} \eta$       & 0.20            & 0.05            \\
 $\Sigma_b^{*+} \eta$      & 0.10            & 0.24            \\
 $\Xi_b^{0} K^+$            & 0.45            & 0.45            \\
 $\Xi_b^{\prime 0} K^+$     & 0.39            & 0.10\\
 $\Xi_b^{\prime *0} K^+$    & 0.17            & 0.46            \\
 $B^0 p$                    & 0.42            & 1.97           \\
 $B^{*0} p$                 & 1.34            & 1.41          \\
 $B^- \Delta^{++}$          & 34.21            & 22.04            \\
 $B^0 \Delta^{+}$           & 11.39            & 7.34           \\
 $B^{*-} \Delta^{++}$          & 28.10         & 50.67         \\
 $B^{*0} \Delta^{+}$           & 9.37         & 16.89             \\
 $B_s \Sigma^{+}$           & 0.22            & 1.33            \\
 Total                      & 87.46           & 104.06          \\
\hline\hline
\end{tabular*}
\end{center}
\end{table}

\subsection{$\Sigma_Q(2P)$}

Five $\lambda$-mode $\Sigma_c(2P)$ states exist theoretically, and the estimated masses in the relativistic quark model are about $3125-3172$ MeV. Until now, there is no experimental information for these states. With the predicted masses within the relativistic quark model~\cite{Ebert:2011kk}, their strong decays are calculated and listed in Tab.~\ref{sigc2p}. The total decay widths of these five states are about $23-100$ MeV, which are relatively narrow. Roughly speaking, the dominant decay modes are light baryon plus heavy meson channels. These relatively narrow total widths and large $D^{(*)}N$ partial decay widths suggest that these states may be easily observed in future experiments.

In previous work~\cite{Lu:2018utx}, the strong decay behaviors have been investigated with the calculated masses in the relativized quark model proposed by Capstick and Isgur~\cite{Capstick:1986bm}. Due to the small differences of the initial baryon masses, the total decay widths change a little, but the conclusions are similar to present ones.

In the same way, the strong decays of five $\Sigma_b(2P)$ states are investigated and presented in Tab.~\ref{sigb2p}. Since the predicted masses of $\Sigma(2P)$ states are lower than the $B\Delta$ threshold, the total widths and dominated decay channels are quite different with the charmed baryon sector. For the $\Sigma_{b0}(\frac{1}{2}^-,2P)$ state, the main decay channel is $\Xi_b K$, while other modes are rather small. For the two $j=1$ states, the dominating decay modes are $\Lambda_b \rho$, $BN$, and $B^*N$ final states. From the heavy quark spin symmetry, the two $j=1$ states should have similar properties, such as masses and total widths. Indeed, the predicted masses and total decay widths of these two states are almost same. The partial decay width ratios of $BN$ and $B^*N$ are predicted to be
\begin{eqnarray}
\Gamma[\Sigma_{b1}\left(\frac{1}{2}^-,2P\right) \to BN]:\Gamma[\Sigma_{b1}\left(\frac{1}{2}^-,2P\right) \to B^*N]  = 0.14
\end{eqnarray}
and
\begin{eqnarray}
\Gamma[\Sigma_{b1}\left(\frac{3}{2}^-,2P\right) \to BN]:\Gamma[\Sigma_{b1}\left(\frac{3}{2}^-,2P\right) \to B^*N]  = 0.65,
\end{eqnarray}
which are essential to distinguish them. The two $j=2$ states mainly decay into the $\Lambda \pi$, $BN$, and $B^*N$ final states, and also have similar masses and total decay widths. The partial decay width ratios of $BN$ and $B^*N$ are
\begin{eqnarray}
\Gamma[\Sigma_{b2}\left(\frac{3}{2}^-,2P\right) \to BN]:\Gamma[\Sigma_{b1}\left(\frac{1}{2}^-,2P\right) \to B^*N]  = 0.03
\end{eqnarray}
and
\begin{eqnarray}
\Gamma[\Sigma_{b2}\left(\frac{5}{2}^-,2P\right) \to BN]:\Gamma[\Sigma_{b1}\left(\frac{3}{2}^-,2P\right) \to B^*N]  = 0.75.
\end{eqnarray}
These five narrow $\Sigma_b(2P)$ states can be searched in $BN$ and $B^*N$ final states, and the ratios of $BN$ and $B^*N$ can help us to distinguish the states with same light spin $j$.

\begin{table}
\begin{center}
\caption{ \label{sigc2p} Decay widths of the $\Sigma_c(2P)$ states in MeV.}
\renewcommand{\arraystretch}{1.5}
\footnotesize
\begin{tabular*}{8.5cm}{@{\extracolsep{\fill}}*{6}{p{1.2cm}<{\centering}}}
\hline\hline
 Mode      & $\Sigma_{c0}(\frac{1}{2}^-,2P)$  & $\Sigma_{c1}(\frac{1}{2}^-,2P)$ & $\Sigma_{c1}(\frac{3}{2}^-,2P)$  &  $\Sigma_{c2}(\frac{3}{2}^-,2P)$ & $\Sigma_{c2}(\frac{5}{2}^-,2P)$ \\
 $\Lambda_c \pi^+$          & 0.91                 & $\cdots$               & $\cdots$             &  2.09              &  2.12 \\
 $\Lambda_c \rho^+$        & $\cdots$                   &  4.42                &  3.60            &  0.31              &  0.39  \\
 $\Sigma_c^{++} \pi^0$     & $\cdots$                   &  0.08                &  0.63            &  1.06              &  0.49\\
 $\Sigma_c^{+} \pi^+$      & $\cdots$                   &  0.08                &  0.63            &  1.06              &  0.49\\
 $\Sigma_c^{*++} \pi^0$     & $\cdots$                    &  0.77                &  0.63            &  0.80              &  1.30\\
 $\Sigma_c^{*+} \pi^+$      & $\cdots$                     &  0.77                &  0.63            &  0.80              &  1.30\\
 $\Sigma_c^{++} \eta$      & $\cdots$                   &  1.50                &  0.09            &  0.12           &  0.06\\
 $\Sigma_c^{*++} \eta$     & $\cdots$                    &  0.02                &  1.67            & 0.03           &  0.07\\
 $\Xi_c^{+} K^+$            & 5.36                  & $\cdots$             & $\cdots$             &  0.51              & 0.58\\
 $\Xi_c^{\prime +} K^+$   & $\cdots$                    &  4.93                &  0.05           &  0.05             &  0.03\\
 $\Xi_c^{\prime *+} K^+$    & $\cdots$                   & $\cdots$               &  4.25            &  $4\times10^{-4}$  &  $3\times10^{-3}$\\
 $D^+ p$                    & 3.74                    &  4.31                &  3.31            & 0.67              &  10.66\\
 $D^{*+} p$                 & 0.39                    &  7.69                &  6.71           &  26.45             &  17.28\\
 $D^0 \Delta^{++}$          & 6.83                    &  0.38               &  14.07            &  49.42              &  3.43\\
 $D^+ \Delta^{+}$           & 2.00                    &  0.08                &  4.60            &  17.26              &  0.98\\
 $D_s \Sigma^{+}$           & 4.07                   & $\cdots$               & 0.01         & $\cdots$  & $4\times10^{-4}$ \\
 Total                      & 23.31                   & 25.03               &  40.88               &  100.62            & 39.18   \\
\hline\hline
\end{tabular*}
\end{center}
\end{table}

\begin{table}
\begin{center}
\caption{ \label{sigb2p} Decay widths of the $\Sigma_b(2P)$ states in MeV.}
\renewcommand{\arraystretch}{1.5}
\footnotesize
\begin{tabular*}{8.5cm}{@{\extracolsep{\fill}}*{6}{p{1.2cm}<{\centering}}}
\hline\hline
 Mode     & $\Sigma_{b0}(\frac{1}{2}^-,2P)$   & $\Sigma_{b1}(\frac{1}{2}^-,2P)$ & $\Sigma_{b1}(\frac{3}{2}^-,2P)$  &  $\Sigma_{b2}(\frac{3}{2}^-,2P)$ & $\Sigma_{b2}(\frac{5}{2}^-,2P)$ \\
 $\Lambda_b \pi^+$          & 0.73                & $\cdots$             & $\cdots$             &  2.84              &  2.83 \\
 $\Lambda_b \rho^+$       & $\cdots$              &  6.44                &  6.44            &  0.04               &  0.03  \\
 $\Sigma_b^{+} \pi^0$     & $\cdots$                &  0.39                &  0.60            & 1.04              &  0.46\\
 $\Sigma_b^{0} \pi^+$     & $\cdots$               &  0.40                &  0.60            &1.04                &  0.46\\
 $\Sigma_b^{*+} \pi^0$    & $\cdots$               &  1.06                &  1.10            &  0.91              &  1.40\\
 $\Sigma_b^{*0} \pi^+$    & $\cdots$               &  1.06                &  1.11            &  0.91              &  1.40\\
 $\Sigma_b^{+} \eta$      & $\cdots$                &  2.51               &  0.02             &  0.03               &  0.01\\
 $\Sigma_b^{*+} \eta$     & $\cdots$                  &  0.02                &  2.47           &  0.01              &  0.01\\
 $\Xi_b^{0} K^+$            & 7.78               & $\cdots$           & $\cdots$           &  0.29            &  0.28\\
 $\Xi_b^{\prime 0} K^+$    & $\cdots$               &  1.07                &  $1\times10^{-6}$    & $\cdots$           & $\cdots$  \\
 $B^0 p$                    & 1.18                & 1.55                &  4.50            &  0.87              &  13.86\\
 $B^{*0} p$                 & 0.06                  &  11.00                &  6.91           &  30.39             &  18.38\\
 Total                      & 9.75             &  25.50               &  23.75            &  38.35           & 39.11   \\
\hline\hline
\end{tabular*}
\end{center}
\end{table}

\subsection{$\Sigma_Q(2D)$}

In the constituent quark model, six $\Sigma_c(2D)$ states are allowed to exist. The predicted masses are much higher than the $DN$, $D^*N$, $D\Delta$, and $D^*\Delta$ thresholds, and these channels provide dominating contributions to their total decay widths. From Tab.~\ref{sigc2d}, the total decay widths of these six states lie in $41-102$ MeV, which can be searched in the heavy meson plus light baryon channels experimentally.

The six $\Sigma_b(2D)$ states have rather higher masses of around 6600 MeV. From Tab.~\ref{sigb2d}, it can be seen that they mainly decay into the heavy meson plus light baryon final states, which are similar to the charmed partners. The total decay widths varies from 44 to 102 MeV, which can be tested in future experiments.

\begin{table*}
\begin{center}
\caption{ \label{sigc2d} Decay widths of the $\Sigma_c(2D)$ states in MeV.}
\renewcommand{\arraystretch}{1.5}
\begin{tabular*}{17cm}{@{\extracolsep{\fill}}*{7}{p{2cm}<{\centering}}}
\hline\hline
 Mode     & $\Sigma_{c1}(\frac{1}{2}^+,2D)$  & $\Sigma_{c1}(\frac{3}{2}^+,2D)$  & $\Sigma_{c2}(\frac{3}{2}^+,2D)$ & $\Sigma_{c2}(\frac{5}{2}^+,2D)$  &  $\Sigma_{c3}(\frac{5}{2}^+,2D)$ & $\Sigma_{c3}(\frac{7}{2}^+,2D)$ \\
 $\Lambda_c \pi^+$          & 0.02            & 0.02         & $\cdots$              & $\cdots$             &  0.29              &  0.28 \\
 $\Lambda_c \rho^+$         & 0.10            & 0.10            &  0.28                &  0.28            &  0.13              &  0.12  \\
 $\Sigma_c^{++} \pi^0$      & $5\times10^{-3}$& $1\times10^{-3}$&  0.01                &  0.11            &  0.12              &  0.06\\
 $\Sigma_c^{+} \pi^+$       &$5\times10^{-3}$ & $1\times10^{-3}$&  0.01                &  0.11            &  0.12              &  0.06\\
 $\Sigma_c^{*++} \pi^0$     & 0.01            & 0.02            &  0.16                &  0.11            &  0.11              &  0.15\\
 $\Sigma_c^{*+} \pi^+$      & 0.01            & 0.02            &  0.16                &  0.11            &  0.11             &  0.15\\
 $\Sigma_c^{++} \rho^0$      & 0.05           & 0.07            &0.15                  &  0.02            &  0.52             & $3\times10^{-3}$ \\
 $\Sigma_c^{+} \rho^+$       & 0.05           & 0.07            &0.15                  &  0.02            &  0.52               &$3\times10^{-3}$ \\
 $\Sigma_c^{*++} \rho^0$     & 0.05           & 0.06             &  0.03                 &   0.10            &  0.07               & 0.28  \\
 $\Sigma_c^{*+} \rho^+$      & 0.06           & 0.06             &  0.03                 &   0.11            &  0.07               &  0.28  \\
 $\Sigma_c^{++} \eta$       & 0.03            & 0.01            &  0.07                &  0.03            &  0.03                &  0.01\\
 $\Sigma_c^{*++} \eta$      & 0.02            & 0.05            &  0.04                &  0.12            & 0.02             &  0.02\\
 $\Sigma_c^{++} \omega$       & 0.05          & 0.06            & 0.14                &  0.02             &  0.51              & $2\times10^{-3}$ \\
 $\Sigma_c^{*++} \omega$      & 0.05          & 0.05              & 0.03                &   0.09            &  0.06              & 0.23  \\
 $\Xi_c^{+} K^+$            & 0.32            & 0.32         & $\cdots$              & $\cdots$            &  0.10              & 0.09\\
 $\Xi_c^{\prime +} K^+$     & 0.13            & 0.03            &  0.29                &  0.02           &  0.02              &  0.01\\
 $\Xi_c^{\prime *+} K^+$    & 0.06            & 0.15            &  0.07                &  0.33            &  0.01              &  0.01\\
 $\Xi_c^{+} K^{*+}$         & 0.01            & $5\times10^{-3}$& $5\times10^{-3}$        & 0.01           & $\cdots$          & $\cdots$  \\
 $D^+ p$                    & 6.38            & 0.39            &  3.43                &  2.01            & 0.15              &  5.46\\
 $D^{*+} p$                 & 0.81            & 2.10           &  8.50                &  5.97           &  19.35             &  11.03\\
 $D^0 \Delta^{++}$          & 0.15           & 19.25            &  2.82               &  15.16            &  3.95              &  7.66\\
 $D^+ \Delta^{+}$           & 0.06            & 6.24            &  0.97                &  4.92            &  1.46              &  2.46\\
 $D^{*0} \Delta^{++}$          & 21.54         & 16.52           &  23.33                &  30.00           &  26.01              &  54.55  \\
 $D^{*+} \Delta^{+}$           & 6.83         & 5.31            &  7.57                &  9.93            & 8.58            &  18.11    \\
 $D_s \Sigma^{+}$           & 1.11            & 0.08            &  0.71                & 0.45          &  0.03  & 0.87 \\
 $D_s^* \Sigma^{+}$           & 2.32          & 6.41            &  1.54                &  4.01             &  0.07   & 0.02 \\
 $D_s \Sigma^{*+}$           & 0.77           & 0.03            &  0.26                & $1\times10^{-4}$ & $\cdots$ & $\cdots$ \\
 Total                      & 40.98           & 57.40           &  50.75               &  74.07            &  62.41    & 101.95   \\
\hline\hline
\end{tabular*}
\end{center}
\end{table*}

\begin{table*}
\begin{center}
\caption{ \label{sigb2d} Decay widths of the $\Sigma_b(2D)$ states in MeV.}
\renewcommand{\arraystretch}{1.5}
\begin{tabular*}{17cm}{@{\extracolsep{\fill}}*{7}{p{2cm}<{\centering}}}
\hline\hline
 Mode     & $\Sigma_{b1}(\frac{1}{2}^+,2D)$  & $\Sigma_{b1}(\frac{3}{2}^+,2D)$  & $\Sigma_{b2}(\frac{3}{2}^+,2D)$ & $\Sigma_{b2}(\frac{5}{2}^+,2D)$  &  $\Sigma_{b3}(\frac{5}{2}^+,2D)$ & $\Sigma_{b3}(\frac{7}{2}^+,2D)$ \\
 $\Lambda_b \pi^+$          & 0.02            & 0.02          & $\cdots$             & $\cdots$            &  0.40              &  0.39 \\
 $\Lambda_b \rho^+$         & 0.19            & 0.18            &  0.47                &  0.47            &  0.10              &  0.09  \\
 $\Sigma_b^{+} \pi^0$      & 0.02            & 0.01            &  0.07                &  0.11            & 0.12                &  0.06\\
 $\Sigma_b^{0} \pi^+$       & 0.02            & 0.01            &  0.08                &  0.11            &0.12                &  0.06\\
 $\Sigma_b^{*+} \pi^0$     & 0.02            & 0.03            &  0.20                &  0.19            &  0.13              &  0.17\\
 $\Sigma_b^{*0} \pi^+$      & 0.02            & 0.04            &  0.20                &  0.19            &  0.13              &  0.17\\
 $\Sigma_b^{+} \rho^0$      & 0.03           &0.03              &0.04                   &  $3\times10^{-3}$& 0.05              &  $3\times10^{-8}$\\
 $\Sigma_b^{0} \rho^+$       &0.03           &0.03             &0.04                    &  $3\times10^{-3}$& 0.05              &  $4\times10^{-8}$\\
 $\Sigma_b^{*+} \rho^0$     & 0.02            & 0.04            &  $1\times10^{-3}$    &$4\times10^{-3}$& $\cdots$            & $\cdots$\\
 $\Sigma_b^{*0} \rho^+$      &0.02           & 0.04            &  $1\times10^{-3}$    &$4\times10^{-3}$ & $\cdots$          & $\cdots$\\
 $\Sigma_b^{+} \eta$       & 0.06            & 0.01            &  0.14                &  0.02           &  0.01               &  $7\times10^{-3}$\\
 $\Sigma_b^{*+} \eta$      & 0.03            & 0.08            &  0.05                &  0.19            &  0.01             &  0.02\\
 $\Sigma_b^{+} \omega$       & 0.03          &0.03             &0.02                  &  $1\times10^{-3}$&  0.01              & $\cdots$\\
 $\Sigma_b^{*+} \omega$      &0.02           &0.03         & $\cdots$                & $\cdots$            & $\cdots$            & $\cdots$\\
 $\Xi_b^{0} K^+$            & 0.39            & 0.38        & $\cdots$             & $\cdots$             &  0.06             &  0.06\\
 $\Xi_b^{\prime 0} K^+$     & 0.13            & 0.03           &  0.26                &  0.01           &  $4\times10^{-3}$  &  $2\times10^{-3}$\\
 $\Xi_b^{\prime *0} K^+$    & 0.06            & 0.16            &  0.06              &  0.30            &  $4\times10^{-3}$  &  $4\times10^{-3}$\\
 $B^0 p$                    & 6.13            & 0.44            &  2.42                &  3.59            &  0.26              &  9.26\\
 $B^{*0} p$                 & 1.78            & 6.37           &  10.76                &  7.62           &  24.45             &  13.89\\
 $B^- \Delta^{++}$          & 3.23            & 14.50            &  14.78               &  7.16            &  34.65              &  1.43\\
 $B^0 \Delta^{+}$           & 1.08            & 4.81            &  4.93                &  2.37            &  11.55              &  0.47\\
 $B^{*-} \Delta^{++}$          & 17.39         & 19.31           &  18.46                &  30.81           &  22.02              &  42.38\\
 $B^{*0} \Delta^{+}$           & 5.80         & 6.44            &  6.15                &  10.27            &  7.34               &  14.13  \\
 $B_s \Sigma^{+}$           & 5.93            & 0.37            &  2.90                &  0.04            &  $1\times10^{-3}$  & 0.02 \\
 $B_s^* \Sigma^{+}$           & 1.96          & 7.56            &  0.19                &  0.51           & $\cdots$           & $\cdots$\\
 Total                      & 44.41           & 60.96           &  62.24               &  63.97            &  101.49    & 82.62   \\
\hline\hline
\end{tabular*}
\end{center}
\end{table*}

\subsection{$\Sigma_Q(1F)$}

For the six $\Sigma_c(1F)$ states, the predicted masses are about $3209-3288$ MeV, which are near the $D^*\Delta$ threshold. With these masses, the calculated strong decay widths are shown in Tab.~\ref{sigc1f}. Roughly speaking, the $DN$, $D^*N$ and $D\Delta$ decay modes are the dominating, while other decay channels seem to be small. Their calculated total decay widths are about $6-53$ MeV, which are relatively narrow. The $DN$, $D^*N$ and $D\Delta$ channels may be ideal channels to hunt for these $F-$wave states.

The predicted masses of six $\Sigma_b(1F)$ states are around 6500 MeV. Due to the limit of phase space, the $B\Delta$ and $B^*\Delta$ decay modes are forbidden or provide negligible contributions. In this situation, the $BN$ and $B^*N$ channels become the dominating decay modes, and the $\Lambda_b \pi$ channel is also important for the $j=2$ and $j=4$ states.  From Tab.~\ref{sigb1f}, the calculated total decay widths range from 4 to 46 MeV, and the ideal channels to search for these states are $BN$, $B^*N$ and $\Lambda_b \pi$ final states.

\begin{table*}
\begin{center}
\caption{ \label{sigc1f} Decay widths of the $\Sigma_c(1F)$ states in MeV.}
\renewcommand{\arraystretch}{1.5}
\begin{tabular*}{17cm}{@{\extracolsep{\fill}}*{7}{p{2cm}<{\centering}}}
\hline\hline
 Mode     & $\Sigma_{c2}(\frac{3}{2}^-,1F)$  & $\Sigma_{c2}(\frac{5}{2}^-,1F)$  & $\Sigma_{c3}(\frac{5}{2}^-,1F)$ & $\Sigma_{c3}(\frac{7}{2}^-,1F)$  &  $\Sigma_{c4}(\frac{7}{2}^-,1F)$ & $\Sigma_{c4}(\frac{9}{2}^-,1F)$ \\
 $\Lambda_c \pi^+$          & 1.60            & 1.60          & $\cdots$              & $\cdots$            &  0.41              &  0.36 \\
 $\Lambda_c \rho^+$         & 0.56            & 0.54            &  0.60                &  0.60            &  0.01              &  $5\times10^{-3}$  \\
 $\Sigma_c^{++} \pi^0$      & 0.46            & 0.20            &  0.94                &  0.08            &  0.06              &  0.03\\
 $\Sigma_c^{+} \pi^+$       & 0.46            & 0.20            &  0.94                &  0.08            &  0.06              &  0.03\\
 $\Sigma_c^{*++} \pi^0$     & 0.39            & 0.60            &  0.30                &  1.03            &  0.04              &  0.04\\
 $\Sigma_c^{*+} \pi^+$      & 0.39            & 0.60            &  0.30                &  1.03            &  0.04              &  0.04\\
 $\Sigma_c^{++} \rho^0$      & 0.01           & 0.01            &$4\times10^{-3}$      &  $4\times10^{-4}$& $\cdots$            & $\cdots$  \\
 $\Sigma_c^{+} \rho^+$       & 0.01           & 0.01            &$4\times10^{-3}$      &  $4\times10^{-4}$& $\cdots$          & $\cdots$  \\
 $\Sigma_c^{++} \eta$       & 0.13            & 0.06            &  0.23                &  $4\times10^{-3}$&  $2\times10^{-3}$  &  $1\times10^{-3}$\\
 $\Sigma_c^{*++} \eta$      & 0.08            & 0.12            &  0.04                &  0.16            & $7\times10^{-4}$ &  $5\times10^{-4}$\\
 $\Sigma_c^{++} \omega$       & 0.01          & $4\times10^{-3}$& $2\times10^{-3}$    &  $2\times10^{-4}$& $\cdots$            & $\cdots$  \\
 $\Xi_c^{+} K^+$            & 0.52            & 0.51          & $\cdots$           & $\cdots$             &  0.01              & 0.01\\
 $\Xi_c^{\prime +} K^+$     & 0.08            & 0.03            &  0.11                &  $7\times10^{-4}$&  $3\times10^{-4}$  &  $1\times10^{-4}$\\
 $\Xi_c^{\prime *+} K^+$    & 0.03            & 0.04            &  0.01                &  0.05            &  $3\times10^{-5}$  &  $1\times10^{-5}$\\
 $D^+ p$                    & 17.60            & 0.49            &  9.94                &  2.55            & 0.08              &  4.12\\
 $D^{*+} p$                 & 14.03            & 34.17           &  10.04                &  19.75           &  3.73             &  1.60\\
 $D^0 \Delta^{++}$          & 8.09           & 2.92            &  15.04               &  1.75            &  20.33              &  0.10\\
 $D^+ \Delta^{+}$           & 2.60            & 0.90            &  4.77                &  0.54            &  6.35              &  0.03\\
 $D^{*0} \Delta^{++}$          & 0.78         & 1.17           &  0.12                &  0.20          & $\cdots$           & $\cdots$  \\
 $D^{*+} \Delta^{+}$           & 0.22         & 0.33            &  0.02                &  0.03         & $\cdots$           & $\cdots$    \\
 $D_s \Sigma^{+}$           & 4.54            & 0.12            &  1.35                & 0.01          &  $5\times10^{-5}$  & $9\times10^{-4}$ \\
 Total                      & 52.61           & 44.62           &  44.76               &  27.86            &  31.12    & 6.37   \\
\hline\hline
\end{tabular*}
\end{center}
\end{table*}

\begin{table*}
\begin{center}
\caption{ \label{sigb1f} Decay widths of the $\Sigma_b(1F)$ states in MeV.}
\renewcommand{\arraystretch}{1.5}
\begin{tabular*}{17cm}{@{\extracolsep{\fill}}*{7}{p{2cm}<{\centering}}}
\hline\hline
 Mode     & $\Sigma_{b2}(\frac{3}{2}^-,1F)$  & $\Sigma_{b2}(\frac{5}{2}^-,1F)$  & $\Sigma_{b3}(\frac{5}{2}^-,1F)$ & $\Sigma_{b3}(\frac{7}{2}^-,1F)$  &  $\Sigma_{b4}(\frac{7}{2}^-,1F)$ & $\Sigma_{b4}(\frac{9}{2}^-,1F)$ \\
 $\Lambda_b \pi^+$          & 1.95            & 1.95          & $\cdots$              & $\cdots$             &  0.37              &  0.33 \\
 $\Lambda_b \rho^+$         & 0.43            & 0.51            &  0.28                &  0.28            &  $6\times10^{-4}$  &  $2\times10^{-4}$  \\
 $\Sigma_b^{+} \pi^0$      & 0.47            & 0.22            &  0.86                &  0.04            & 0.03                &  0.01\\
 $\Sigma_b^{0} \pi^+$       & 0.47            & 0.22            &  0.86                &  0.04            &0.03                 &  0.01\\
 $\Sigma_b^{*+} \pi^0$     & 0.43            & 0.71            &  0.28                &  1.02            &  0.03              &  0.03\\
 $\Sigma_b^{*0} \pi^+$      & 0.43            & 0.71            &  0.28                &  1.02            &  0.03              &  0.03\\
 $\Sigma_b^{+} \eta$       & 0.08            & 0.04            &  0.09                &  $5\times10^{-4}$&  $1\times10^{-4}$  &  $5\times10^{-5}$\\
 $\Sigma_b^{*+} \eta$      & 0.06            & 0.12            &  0.02                &  0.08            &  $8\times10^{-5}$   &  $5\times10^{-5}$\\
 $\Xi_b^{0} K^+$            & 0.46            & 0.51           & $\cdots$           & $\cdots$             &  $2\times10^{-3}$  &  $1\times10^{-3}$\\
 $\Xi_b^{\prime 0} K^+$     & 0.02            & 0.01          &  0.01                &  $1\times10^{-5}$ &  $9\times10^{-7}$  &  $1\times10^{-7}$\\
 $\Xi_b^{\prime *0} K^+$    & 0.02            & 0.04            &  $2\times10^{-3}$    &  0.01            &  $2\times10^{-7}$  &  $1\times10^{-8}$\\
 $B^0 p$                    & 23.36            & 0.65            &  12.25                &  1.51            &  0.04              &  1.94\\
 $B^{*0} p$                 & 16.57            & 41.53           &  10.75                &  21.72           &  2.99             &  1.29\\
 $B^- \Delta^{++}$          & 0.87            & 0.26            & $\cdots$              & $\cdots$           & $\cdots$             & $\cdots$\\
 $B^0 \Delta^{+}$           & 0.29            & 0.09            & $\cdots$                & $\cdots$           & $\cdots$            & $\cdots$\\
 $B^{*-} \Delta^{++}$       & $\cdots$         & 0.04           & $\cdots$                & $\cdots$          & $\cdots$             & $\cdots$\\
 $B^{*0} \Delta^{+}$        & $\cdots$          & 0.01            & $\cdots$               & $\cdots$           & $\cdots$             & $\cdots$  \\
 $B_s \Sigma^{+}$           & $\cdots$    & $5\times10^{-4}$    & $\cdots$               & $\cdots$           & $\cdots$             & $\cdots$ \\
 Total                      & 45.91           & 47.62           &  25.68               &  25.72            &  3.51    & 3.65   \\
\hline\hline
\end{tabular*}
\end{center}
\end{table*}
\subsection{Discussions on decay modes}

From our present calculations, it can be found that most of the $\lambda-$mode higher excited $\Lambda_Q$ and $\Sigma_Q$ states mainly decay into the final states with one heavy meson plus one light baryon. Several factors, such as phase space, flavor overlap, Clebsch-Gordan coefficients, and spatial overlaps, may result in this special feature. After careful examinations, it is found that some significant partial decay widths of heavy meson plus light baryon modes in various places mainly come from the larger spatial overlaps, while the phase space, flavor overlap, and Clebsch-Gordan coefficients may cause secondary contribution.

\begin{figure}[htb]
\includegraphics[scale=0.75]{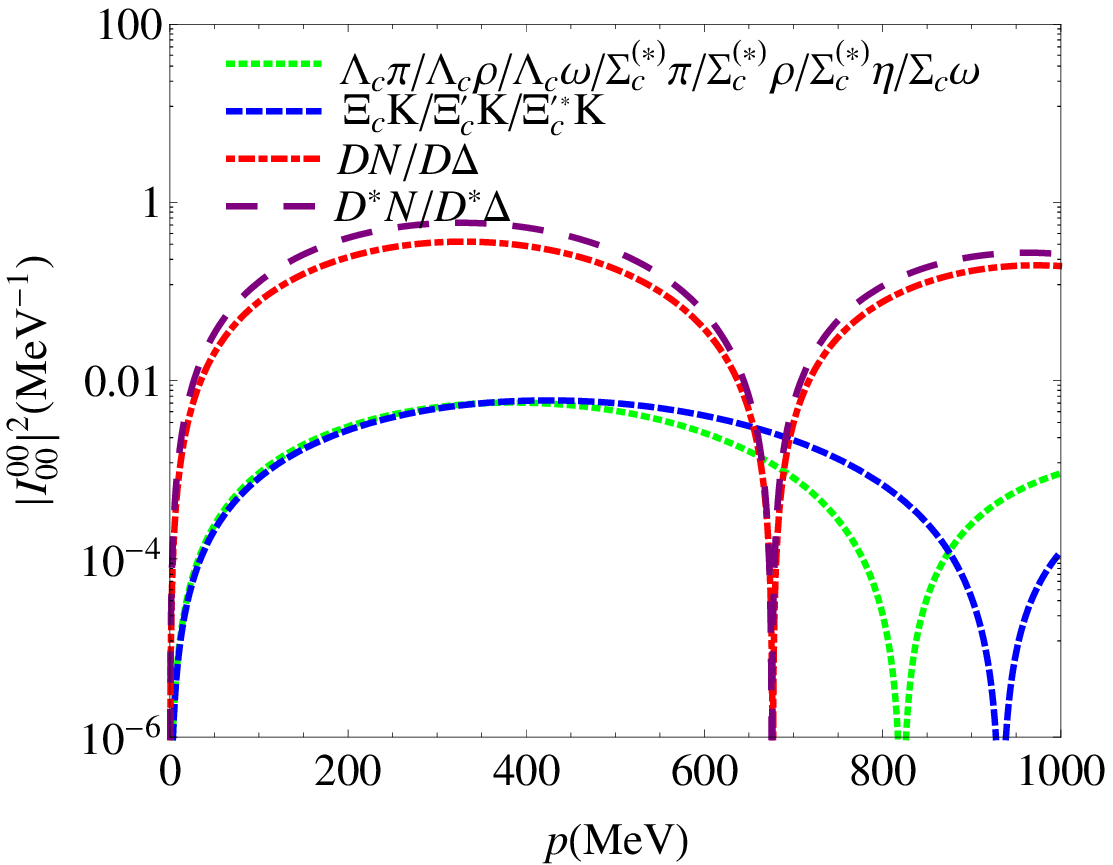}
\vspace{0.0cm} \caption{The spatial overlaps $|I^{00}_{00}(p)|^2$ versus the final momentum $p$ for the initial $\Lambda_c(3S)$, $\Sigma_c(3S)$, and $\Sigma_c^*(3S)$ baryons.}
\label{I3S}
\end{figure}

For instance, we plot the spatial overlaps $|I^{M_{L_A}m}_{M_{L_B}M_{L_C}}(p)|^2$ versus the final momentum $p$ for the initial $\Lambda_c(3S)$, $\Sigma_c(3S)$, and $\Sigma_c^*(3S)$ baryons in Fig.~\ref{I3S}. In this case, only the $I^{00}_{00}(p)$s are nonzero, and four types of spatial overlaps can be classified by the final states. It can be seen that for the heavy meson plus light baryon modes the spatial overlaps are significantly larger in most regions. For some cases, together with the effects of phase space, flavor overlap, and Clebsch-Gordan coefficients, the partial decay widths of heavy meson plus light baryon modes are larger than that of heavy baryon plus light meson channels due to the significant spatial overlaps.

One possible interpretation of the significant spatial overlaps of some heavy meson plus light baryon modes is given as follows. For the higher excited $\lambda-$mode states, the light quark subsystem has no excitation and the final states also lie in the ground states. When the heavy meson plus light baryon decay mode occurs, the whole light quark subsystem goes into the final baryon as a spectator. Without the extra separation, the overlap of $\rho-$mode wavefunctions between initial and final baryons equals to 100\% for the heavy meson plus light baryon mode, and then the total spatial overlap may be large enough.

\section{Summary}{\label{Summary}}

 In this work, we preform a systematic study of the strong decays of the higher excited singly heavy baryons $\Lambda_Q$ and $\Sigma_Q$ states. With the predicted masses of the relativistic quark model, the two body OZI-allowed strong decay widths of $\Lambda_Q(3S)$, $\Lambda_Q(2P)$, $\Lambda_Q(2D)$, $\Lambda_Q(1F)$, $\Sigma_Q(3S)$, $\Sigma_Q(2P)$, $\Sigma_Q(2D)$, and $\Sigma_Q(1F)$ states are calculated by using the $^3P_0$ quark pair creation model. Our results may provide useful information to look for these higher $\Lambda_Q$ and $\Sigma_Q$ states.

It should be mention that most of the $\lambda-$mode higher excited $\Lambda_Q$ and $\Sigma_Q$ states have relatively narrow total widths, and mainly decay into the heavy meson plus light baryon final states rather than heavy baryon plus light meson decay modes. For the heavy meson plus light baryon case, the whole light quark subsystem goes into the final baryon as a spectator and the overlap of $\rho-$mode wavefunctions between initial and final states equals to 100\%, which might make the strong decay easier. These specific decay modes can help us to establish the $\lambda-$mode higher singly baryon spectrum since the $\rho-$mode states decaying into the heavy meson plus light baryon channels are highly suppressed. To be more specific, most of the higher $\Lambda_Q$ states mainly decay into the $D^{(*)}N$ or $B^{(*)}N$ channels, while most of the higher $\Sigma_Q$ states mainly decay into the $D^{(*)}\Delta$ or $B^{(*)}\Delta$ final states. However, for some of the $2P$ and $1F$ states, the predicted masses may lie near or below the relevant $D^{(*)}N$, $B^{(*)}N$, $D^{(*)}\Delta$ or $B^{(*)}\Delta$ threshold, and other decay channels may be important due to the phase space constraints.

 Considering the uncertainties of the adopted masses and the $^3P_0$ model, one do not expect the estimated decay widths are accurate. Our predictions should bear uncertainties about $30\%$ and be regarded as a semi-quantitative estimation. Still, our calculations demonstrate the general feature and provide abundant theoretical information for these higher excited singly heavy baryons, which may be helpful for future experimental searches.

\bigskip
\noindent
\begin{center}
{\bf ACKNOWLEDGEMENTS}\\

\end{center}
This project is supported by the National Natural Science Foundation of China under Grants No. 11705056, No.~11775078, and No.~U1832173.

\end{document}